%% file: main.tex
\documentclass{article}

\usepackage{silence}
\PassOptionsToPackage{dvipsnames,table}{xcolor}
\PassOptionsToPackage{numbers, compress}{natbib}

\usepackage[preprint]{neurips_2026}

\usepackage[utf8]{inputenc}
\usepackage[T1]{fontenc}
\usepackage{hyperref}
\usepackage{url}
\usepackage{microtype}
\usepackage{nicefrac}

\input{math_commands.tex}
\input{preamble/preamble.tex}

\usepackage[capitalize,noabbrev]{cleveref}

\title{Parallel Test-Time Scaling with Multi-Sequence Verifiers}

\author{%
  Yegon Kim \\
  Graduate School of AI, KAIST \\
  Seoul, South Korea \\
  \texttt{yegonkim@kaist.ac.kr} \\
  \And
  Seungyoo Lee \\
  Graduate School of AI, KAIST \\
  Seoul, South Korea \\
  \And
  Chaeyun Jang \\
  Graduate School of AI, KAIST \\
  Seoul, South Korea \\
  \And
  Hyungi Lee \\
  Graduate School of AI, KAIST \\
  Seoul, South Korea \\
  \And
  Juho Lee \\
  Graduate School of AI, KAIST \\
  Seoul, South Korea
}

\begin{document}

\maketitle

\input{main/abstract}
\input{main/introduction}
\input{main/relatedwork}
\input{main/method}
\input{main/experiments}

\input{main/conclusion}

\clearpage
\newpage

\begin{ack}
\end{ack}

\bibliography{neurips2026_conference}
\bibliographystyle{plainnat}

\newpage
\appendix

\input{appendix/experimental_details}
\clearpage
\input{appendix/additional_exp}
\input{appendix/method_details}
\clearpage
\input{appendix/full_report}


\end{document}

%% file: math_commands.tex
\usepackage{amsmath,amsfonts,bm}

\def\eqref#1{equation~\ref{#1}}

\def\1{\bm{1}}

\DeclareMathAlphabet{\mathsfit}{\encodingdefault}{\sfdefault}{m}{sl}
\SetMathAlphabet{\mathsfit}{bold}{\encodingdefault}{\sfdefault}{bx}{n}

\DeclareMathOperator*{\argmax}{arg\,max}

\usepackage{scalerel}
\newcommand{\smallth}[1]{^{\scaleto{(#1)}{5.5pt}}}
\newcommand{\nth}{\smallth{n}}
\newcommand{\tth}{\smallth{t}}

%% file: preamble/preamble.tex
\usepackage[most]{tcolorbox}

\tcbset{
  highlight token/.style={
    boxrule=0.4pt,
    colback=gray!5,
    colframe=gray!50,
    fontupper=\ttfamily\color{purple},
    arc=1pt,
    boxsep=1pt,
    left=0.5pt,
    right=0.5pt,
    top=0.5pt,
    bottom=0.5pt,
    enhanced,
    box align=base,
    on line,
    height=1.2em,
    valign=center, 
  }
}

\usepackage[dvipsnames,table]{xcolor}
\definecolor{darkpink}{RGB}{199,21,140}

\usepackage{adjustbox}
\usepackage{multirow}
\usepackage{booktabs, array}
\usepackage{makecell}
\usepackage{tabularx}
\usepackage{wrapfig}
\usepackage{makecell}
\usepackage{enumitem}

\usepackage{graphicx}
\usepackage[labelfont=bf]{caption}
\usepackage[format=hang]{subcaption}

\definecolor{citecolor}{RGB}{0,102,204}
\definecolor{linkcolor}{RGB}{190,105,30}
\definecolor{urlcolor}{RGB}{199,21,133}

\usepackage[acronym,nowarn,section,nogroupskip,nonumberlist]{glossaries}
\glsdisablehyper{}

\usepackage{algorithm,algorithmicx,algpseudocode}

\input{preamble/acronyms.tex}

\input{preamble/math.tex}

\usepackage{caption}

%% file: preamble/acronyms.tex
\newacronym{msv}{\textsc{MSV}}{Multi-Sequence Verifier}

%% file: preamble/math.tex
\newsavebox\CBox

\def\PM#1{\scriptsize{\textpm#1}}

%% file: main/abstract.tex
\begin{abstract}

Parallel test-time scaling, which generates multiple candidate solutions for a single problem, is a powerful technique for improving large language model performance. However, it is hindered by two key bottlenecks: accurately selecting the correct solution from the candidate pool, and the high inference latency from generating many full solutions. We argue that both challenges are fundamentally linked to verifier calibration, as
a well-calibrated verifier improves answer selection and enables early-stopping strategies to reduce latency.
However, existing non-generative verifiers are limited as they score each candidate in isolation, overlooking rich contextual information across the set of candidates.
To address this, we introduce the \gls{msv},
a lightweight verifier that predicts each candidate's correctness conditioned on the full sampled set.
MSV achieves improved calibration, which directly enhances best-of-N selection performance and empowers a novel early-stopping framework.
Across challenging mathematical reasoning benchmarks, MSV improves best-of-64 accuracy by up to 6\% relative to strong baselines, and in the early-stopping setting reaches the same accuracy as baselines with less than half the latency.

\end{abstract}

%% file: main/introduction.tex

\section{Introduction}
\label{sec:intro}


Large language models \citep{brown2020language} have become increasingly powerful, with much of their performance unlocked by test-time scaling strategies. One of the most effective strategies is parallel scaling, where a model generates multiple, independent candidate solutions for a single problem \citep{wang2022self,lightman2023let,snell2024scaling}.
However, parallel scaling faces two bottlenecks that limit its usage: (1) the selection problem, as accurately identifying the correct solution from a large pool of candidates is difficult, and (2) the high inference latency required to generate numerous full solutions.

We argue that these two bottlenecks are not independent;
their solutions can be jointly addressed through the principle of calibration.
The selection problem is fundamentally a classification task, where a verifier \citep{cobbe2021training,lightman2023let} must accurately estimate the correctness of each solution.
Its ability to do so, its calibration \citep{guo2017calibration,kadavath2022language}, affects the performance of downstream parallel scaling methods such as best-of-N. Concurrently, the high cost of generation can be mitigated by early stopping, a technique that also hinges on a well-calibrated verifier to score intermediate answers and terminate decoding once a threshold score is exceeded \citep{zhang2025reasoning, yang2025dynamic}.
Thus, a verifier with superior calibration offers a unified path to addressing both the accuracy and efficiency challenges of parallel scaling.

An underutilized source of signal for verifier calibration is the structure of the language model's sampling distribution. Generating multiple candidate solutions for the same problem does more than produce a pool of answers---it reveals the model's conditional reasoning distribution over that problem. The geometry of this distribution carries meaningful information: candidates whose final answers cluster together or whose reasoning traces follow similar paths provide evidence about the model's confidence and the likely correctness of any individual candidate.
A verifier blind to this structure risks discarding precisely the signal that the sampling process was designed to produce.

Existing methods partially exploit this idea.
For instance, self-consistency \citep{wang2022self} and weighted voting \citep[WV;][]{li2022making} use answer agreement as a simple cross-sequence statistic to score the sampled solutions.
However, these methods face two fundamental shortcomings.
First, answer agreement is a lossy statistic: it ignores the reasoning traces and hidden representations that produced those answers.
Second, they are not optimized end-to-end, leading to potential brittleness.
One salient case of failure occurs when the equivalence checker deviates from the oracle, fragmenting a correct equivalence class or merging incorrect answers with correct ones. In \cref{sec:checker_robustness}, we find that both methods degrade significantly in performance as this deviation becomes larger.
Thus, while they demonstrate the value of cross-sequence information, they also motivate a more robust verifier that can learn rich interactions across candidates.

To address this gap, we introduce the Multi-Sequence Verifier (MSV), a non-generative verifier that learns to use rich cross-sequence information through {end-to-end training}. Instead of estimating correctness from a single sequence alone, MSV conditions each prediction on the hidden representations of all sampled candidates.
Our experiments show that MSV achieves superior calibration compared to strong baselines, and that this improvement helps downstream parallel scaling methods.
First, it enhances best-of-N decoding, leading to a more accurate final answer selection and more reliable confidence scores for the chosen answers.
On challenging math reasoning benchmarks, MSV improves best-of-64 accuracy by up to 6\% relative to baselines. 
Second, and more significantly, we introduce a streaming variant of \gls{msv} that empowers a novel framework for early stopping coupled with parallel decoding.
In this framework, streaming MSV achieves the same peak accuracy as baseline verifier models with less than half the latency.

Our contributions can be summarized as follows:
\begin{itemize}
\item We propose the Multi-Sequence Verifier (MSV), a novel verifier architecture that models cross-sequence interactions to achieve improved calibration.
\item We demonstrate that improved calibration from MSV directly translates to enhanced best-of-N answer selection and more calibrated confidence levels associated with the chosen answers.
\item We generalize an existing early-stopping framework to the parallel decoding setting, and introduce a streaming MSV variant that outperforms strong baselines in this setting.
\end{itemize}


%% file: main/relatedwork.tex
\section{Related Work}
\label{sec:related}

\vspace{-5pt}

\paragraph{Calibration of LLM outputs.}
Confidence estimation for LLMs has been studied through black-box and white-box approaches.
Black-box approaches have focused on prompting a language model to verbalize its confidence \citep{lin2022teaching,tian2023just}. \citet{xiong2023can} find that white-box approaches to calibration, which instead probe internal states \citep{kadavath2022language}, generally outperform their black-box counterparts.
\citet{zhang2025reasoning} demonstrate an interesting result that intermediate answers from reasoning models can be robustly calibrated through a lightweight probe on their hidden states, and apply this finding to early stopping in the decoding of a single sequence.

\vspace{-5pt}

\paragraph{Parallel test-time scaling.}
Answer selection among multiple decoded sequences is most commonly approached by training calibrated verifier models \citep{cobbe2021training,zhang2024generative}.
Additionally, self-consistency \citep{wang2022self} and weighted voting \citep{li2022making} show that a simple global statistic of vote counting across sequences can improve predictions, motivating a broader exploration of cross-sequence signals for selection.
Several \emph{generative} approaches leverage this intuition directly \citep{chen2023universal,liu2025pairjudge,toshniwal2025genselect}. For example, GenSelect \citep{toshniwal2025genselect} prompts an LLM to directly select the best solution from $N$ candidates. While these share MSV's intuition, generative verification incurs inference latencies orders of magnitude higher than MSV's single transformer layer (GenSelect: ${\sim}130$s vs.\ MSV$_{16}$: $0.006$s for $N{=}16$), making it inapplicable in the early stopping setting where a judgment is needed at \emph{every} intermediate answer.
We also provide experiments with GenSelect in \cref{app:genselect}.

\vspace{-5pt}

\paragraph{Adaptive scaling for efficiency.}
The efficiency of parallel test-time scaling is essential for its practical usage.
Various works such as DeepConf \citep{fu2025deep} and others \citep{li2024escape,aggarwal2023let,xue2023dynamic,huang2025efficient} propose early stopping at the sequence level to reduce the number of decoded sequences. However, they assume that multiple sequences are decoded sequentially, while in practice we would be decoding the sequences in parallel.
We thus focus on the parallel decoding setting, with token-level early stopping.
Although token-level early stopping has been explored in the single-sequence setting \citep{zhang2025reasoning, yang2025dynamic},
we are not aware of prior work that studies token-level early stopping in the parallel decoding setting.

\vspace{-5pt}

%% file: main/method.tex
\input{figure/illustration/illustration}

\section{Method}
\label{sec:method}


\subsection{Problem Setup}
\label{subsec:problem_setup}

\vspace{-5pt}

In this paper, we assume a \textit{parallel decoding} or \textit{parallel sampling} scenario, where our LM generates $N$ sequences simultaneously for a given query $q$. At each decoding step, one token is sampled for every sequence, in parallel. This parallel approach reduces latency compared to sequential decoding, a crucial advantage for practical applications.
Under this scenario, we consider two settings: (1) \textbf{Terminal Answers} and (2) \textbf{Streaming Answers}. In brief, the \textbf{Terminal Answers} setting focuses on the final answers obtained from each sequence after decoding is complete. The \textbf{Streaming Answers} setting additionally focuses on the intermediate answers produced while decoding. We describe each setting in more detail below.

\vspace{-5pt}

\paragraph{Terminal Answers.}
Given a query $q$, the LLM decodes $N$ parallel sequences until they all reach termination. To extract an explicit answer, we append an elicitation prompt such as
\verb|### Final Answer ### \boxed| 
to the end of each $n$th sequence, prompting the model to produce a boxed answer $a\nth$. This results in one answer per sequence, denoted by $\{a\nth \}_{n=1}^N$.

\vspace{-5pt}

\paragraph{Streaming Answers.}
Unlike the \textbf{Terminal Answers} setting, we additionally extract \textit{intermediate answers} whenever a delimiter, e.g., the token ``Wait'', is encountered. This approach of extracting intermediate answers at delimiters has been explored in prior work for single-sequence early stopping \citep{zhang2025reasoning, yang2025dynamic}, which we generalize to the parallel decoding setting. Specifically, when the $k$th delimiter appears in the $n$th sequence, we immediately branch that sequence, append the elicitation prompt \verb|### Final Answer ### \boxed|, and obtain an intermediate answer $a_k\nth$. We also extract the terminal answer at the end of each sequence. Thus, if a sequence contains $K\nth-1$ delimiters, it may yield multiple answers $\{a_k\nth\}_{k=1}^{K\nth}$, where
$a_{K\nth}\nth$ is the terminal answer.
For notational consistency, in the \textbf{Terminal Answers} setting, we regard the single terminal answer as $a_1\nth$ and $K\nth = 1$.

Along with each answer, we store the representations of the answer tokens for later use with our Multi-Sequence Verifier. Specifically, if $a_k\nth$ consists of $L_k\nth$ tokens, we store $\{h_{k,i}\nth\}$ for $i=1,\dots, L_k\nth$, where $h_{k,i}\nth \in \mathbb{R}^d$ denotes the $d$-dimensional hidden state of the $i$th token, output by the last transformer layer of the LLM.


\paragraph{Correctness and objective.}
Let $a^*$ denote the ground-truth answer for the query $q$, and let the symbol $\sim$ be an equivalence relation that captures symbolic or semantic equality, e.g., SymPy~\citep{meurer2017sympy,lewkowycz2022solving}. Then, we define the correctness of each candidate answer $a_k\nth$ as follows:
\[
y_k\nth = \mathbf{1}\left[a_k\nth \sim a^*\right].
\]
Our goal is to accurately predict $y_k\nth$ for every candidate, so that we can make more effective use of the generated answers in downstream parallel test-time scaling methods such as best-of-N.

\vspace{-5pt}


\subsection{Multi-Sequence Verifier}
\label{subsec:msv}

\vspace{-5pt}

In this section, we present MSV, our novel verifier architecture that predicts the correctness of each answer by consuming its last-layer hidden states, along with the hidden states of all the other decoded answers. See \cref{fig:mask_schematic} for a schematic diagram of \gls{msv}, and \cref{alg:msv} for a pseudocode of MSV's forward pass.

\vspace{-5pt}

\paragraph{Input representation for \gls{msv}.}
Let $t$ be a time index shared across parallel sequences, where all the sequences start generating from $t=0$ and they generate one token at a time. Let $\tau\nth_k$ denote the time stamp at which the $n$th sequence generates the $k$th answer $a\nth_k$. At a specific readout time step $t$, \gls{msv} takes the representations of all the answers generated up to the step $t$ to predict the correctness of the answers. Specifically, we collect the representations as follows,
\[
U\tth = \operatorname{concat}
\bigg(
\Big\{
\Big\{
\big\{h\nth_{k,i} + \mathsf{e}\nth
\big\}_{i=1}^{L_k\nth}
\Big\}_{\tau_k\nth + L_k\nth \leq t}
\Big\}_{n=1}^N
\bigg).
\]
Here, $\mathrm{concat}(\cdot)$ denotes concatenation of all the representations along the sequence dimension, {without adding any special separators between sequences.} $\mathsf{e}\nth$ is a learnable per-sequence embedding added to identify the sequence from which each token representation originates.

\vspace{-5pt}

\paragraph{Multi-mask transformer blocks.}
To process the aggregated representation $U\tth$, we introduce a \textbf{M}ulti-\textbf{M}ask \textbf{T}ransformer \textbf{B}lock (MMTB). This block captures diverse aspects of the token sequences by combining multiple attention outputs, each derived from a different mask applied to the same input, $U\tth$. As in standard multi-head attention~\citep{vaswani2017attention}, we begin by computing the usual linear projections of $U\tth$ into query, key, and value matrices $(Q\tth_h, K\tth_h, V\tth_h)_{h=1}^H$ where $H$ is the number of heads. For a fixed collection of $J$ masks $\{M_j\}_{j=1}^J$, we then compute the output for each mask:
\[
A\tth_{h,j} \;=\; \operatorname{softmax}\!\left(\frac{Q\tth_h(K\tth_h)^\top}{\sqrt{d_{\mathrm{head}}}} \;+\; \log M_j\right)V_h\tth,
\]
where $\log M_j$ adds $0$ to permitted entries and $-\infty$ to masked entries, with $\operatorname{softmax}$ applied row-wise. The masked outputs are then combined via learnable mixture weights $\mathbf{w}_h=(w_{h,1},\dots,w_{h,J})$ for each head $h$:
\[
\tilde{U}\tth \;=\; \sum_{h=1}^H \sum_{j=1}^J \alpha_{h,j} A\tth_{h,j},
\quad
\bm{\alpha}_h \;=\; \operatorname{softmax}(\mathbf{w}_h).
\]

The collection of masks $\{M_j\}_{j=1}^J$ is a fixed hyperparameter of our model. In our instantiation, we use four complementary masks.
The \emph{full} mask permits all interactions, $(M_{\mathrm{full}})_{u,v}=1$ for all positions $u,v$, enabling attention across all tokens in all sequences.
The \emph{within-sequence} mask restricts attention to tokens originating from the same sequence,
$$(M_{\mathrm{ws}})_{u,v} \;=\; \mathbf{1}\!\big[\mathrm{seq}(u)=\mathrm{seq}(v)\big],$$
capturing within-sequence signals while blocking cross-sequence signals. Here, $\mathrm{seq}(u)$ denotes the index $n$ of the sequence containing the token $u$.
The \emph{equivalence} mask allows attention only between tokens whose answers are symbolically equivalent,
$$(M_{\mathrm{eq}})_{u,v} \;=\; \mathbf{1}\!\big[\mathrm{ans}(u)\ \sim\ \mathrm{ans}(v)\big],$$
where $\mathrm{ans}(u)$ is the answer $a\nth_k$ containing the token $u$.
Finally, the \emph{within-answer} mask allows attention only between tokens inside a single answer instance $a\nth_k$,
$$(M_{\mathrm{wa}})_{u,v} \;=\; \mathbf{1}\!\big[(\mathrm{seq}(u),\mathrm{step}(u))=(\mathrm{seq}(v),\mathrm{step}(v)) \big],$$
where $\mathrm{step}(u)$ identifies the answer step $k$ to which the token belongs.

{
In the \textbf{Terminal Answers} setting, the within-sequence mask and within-answer mask are equivalent, reducing the number of masks to three.
}
We further restrict the attention masks to be ``causal'' in the \textbf{Streaming Answers} setting, meaning that an answer $a\nth_k$ may attend to $a\smallth{n'}_{k'}$ only if $\tau\nth_k + L_{k}\smallth{n} \geq \tau\smallth{n'}_{k'} + L_{k'}\smallth{n'}$.
We provide an ablation study for each mask, in \cref{sec:mask_ablation}.

The block's final output is then computed using standard residual connections and an MLP layer:
$$
Z\tth \;=\; (U\tth + \tilde{U}\tth) + \operatorname{MLP}\!\big(\operatorname{LN}(U\tth + \tilde{U}\tth)\big),
$$

\paragraph{Feature augmentation.}
To provide an explicit agreement signal, we compute the fraction $\gamma_k\nth\in[0,1]$ of sequences whose latest completed answer is symbolically equivalent to $a_k\nth$. We project it through a small MLP which is then added to the hidden states.
Specifically, from $Z\tth$, we extract the representation $z_{k,L_k}\nth$ corresponding to the last token of $a_k\nth$, and add the information from $\gamma_k\nth$, as follows:
\[
\bar{z}_k\nth = z\nth_{k,L_k\nth} + \operatorname{MLP}(\gamma_k\nth).
\]
{We only use the last token representation $z\nth_{k,L_k\nth}$ for computing the prediction, because the attention layer in MMTB has already aggregated the relevant information from all the other tokens in $a_k\nth$.}

\paragraph{Constructing final predictions.}
To predict the final correctness of $a_k\nth$, we apply a linear head to its augmented representation and pass the resulting logit through a sigmoid:
\[
\tilde{y}_{k}\nth \;=\; \sigma\!\big(\mathbf{w}^\top \bar z_{k}\nth + b\big),
\]
where $\sigma$ denotes the sigmoid function and $\mathbf{w},b$ are learnable parameters. Additionally, in the \textbf{Terminal Answers} setting, instead of computing each $\tilde{y}_{1}\nth$ independently, we first average the \emph{logits} of symbolically equivalent terminal answers and then apply the sigmoid function as follows,
\begin{equation}
\label{eq:logit_avg}
\tilde{y}_{1}\nth = \sigma\left( 
\frac{1}{|\mathcal{C}(n)|}
\sum_{m \in \mathcal{C}(n)}
\left(
\mathbf{w}^\top \bar{z}_{1}\smallth{m} + b
\right)
\right)
\end{equation}
where $\mathcal{C}(n) := \{ m \in \{1,\dots, N\}\,|\, a_1\smallth{m} \sim a_1\nth \}$.
Logit averaging incorporates our a priori knowledge that symbolically equivalent answers share the same correctness label.
Logit averaging is inappropriate in the case of \textbf{Streaming Answers}, because the information carried by $\bar{z}_{k}\nth$ grows with $k$, and logits at earlier steps might disrupt rather than complement the prediction at the current step $k$.

\paragraph{Training and Inference.}
For the \textbf{Streaming Answers} scenario, we run the sequences until all of them terminate, collect all the intermediate and final answers produced, and train the parameters to minimize the binary cross-entropy loss. Let $T$ be the global time step when all the sequences terminate. We first compute $Z\smallth{T}$, compute the predictions $\tilde{y}_k\nth$, and minimize
\[
\mathcal{L} := \sum_{n,k} \operatorname{BCE}(\tilde{y}_k\nth, y_k\nth),
\]
where the summation is over all valid $n,k$ pairs.
For the \textbf{Terminal Answers} scenario, we proceed similarly, except that we minimize $\operatorname{BCE}$ only for the terminal answers. {$\operatorname{BCE}$ is a strictly proper loss, which incentivizes calibrated probabilities under standard assumptions. \citep{blasiok2023does}}

At inference time, whenever we want to predict whether the sequences are producing correct answers at some specific time $t$, we compute $Z\tth$, and predict $\{\{\tilde{y}\nth_k\}_{\tau_k\nth + L_{k}\nth\leq t}\}_{n=1}^N$.
The causal attention structure in Streaming Answers enables an efficient online implementation via KV caching, described in \cref{app:online_inference}.


\subsection{Applications}
\label{subsec:applications}
In this section, we present how to use the \gls{msv}-based correctness predictions $\tilde y\nth_k$ in the \textbf{Terminal Answers} and \textbf{Streaming Answers} settings. Basically, in the \textbf{Terminal Answers} setting, we use $\tilde y\nth_1$ as a verifier score for \emph{best-of-N} selection, whereas in the \textbf{Streaming Answers} setting, we use $\tilde y\nth_k$ both as an early-stopping signal \emph{and} as a criterion for selecting the best candidate.

\paragraph{Calibrated best-of-N predictions.}
In the \textbf{Terminal Answers} scenario, our verifier scores are used for \emph{best-of-N} decoding, a robust technique for improving performance~\citep{nakano2021webgpt}. From $N$ candidate answers, we select the one, $a^\dagger$, with the highest score:
$$n^\dagger = \argmax_{n} \tilde y_1\nth, \qquad a^\dagger = a_1\smallth{n^\dagger}.$$

The score for the chosen answer, $\tilde y^\dagger = \tilde y_1\smallth{n^\dagger}$, can serve as a confidence estimate of $a^\dagger$ being correct. We verify empirically that this post-selection score is well-calibrated (see \cref{fig:bon_calibration}), allowing us to assess not only \emph{which} answer is best but also \emph{how likely} it is to be correct.


\paragraph{Early stopping of parallel decoding.}
In the \textbf{Streaming Answers} setting, multiple sequences are decoded in parallel, and each intermediate answer $a\nth_k$ yields a correctness score $\tilde y\nth_k$. We define an early-stopping rule that halts decoding once the best available candidate exceeds a threshold $\lambda$:
\[
t^* \;=\; \min\!\left\{\,t:\; \max_{n,k:\,\tau_k\nth + L_k\nth \le t} \; \tilde y_k\smallth{n} \;\ge\; \lambda \right\}.
\]
At $t^*$, we output the candidate with the highest predicted score, which is the candidate that caused the early-stopping.
The threshold $\lambda \in [0,1]$ is a hyperparameter that can be selected via cross-validation to target a desired operating point on the accuracy-latency tradeoff.
If the system terminates without any of the decoded answers exceeding the threshold, the system outputs the best among the final answers $\tilde y\nth_{K\nth}$.
Thus, ${\tilde y\nth_k}$ acts both as the stopping signal and as verifier scores for selecting the best answer.

%% file: figure/illustration/illustration.tex
\begin{figure*}[t]
    \centering
    \includegraphics[width=0.9\linewidth]{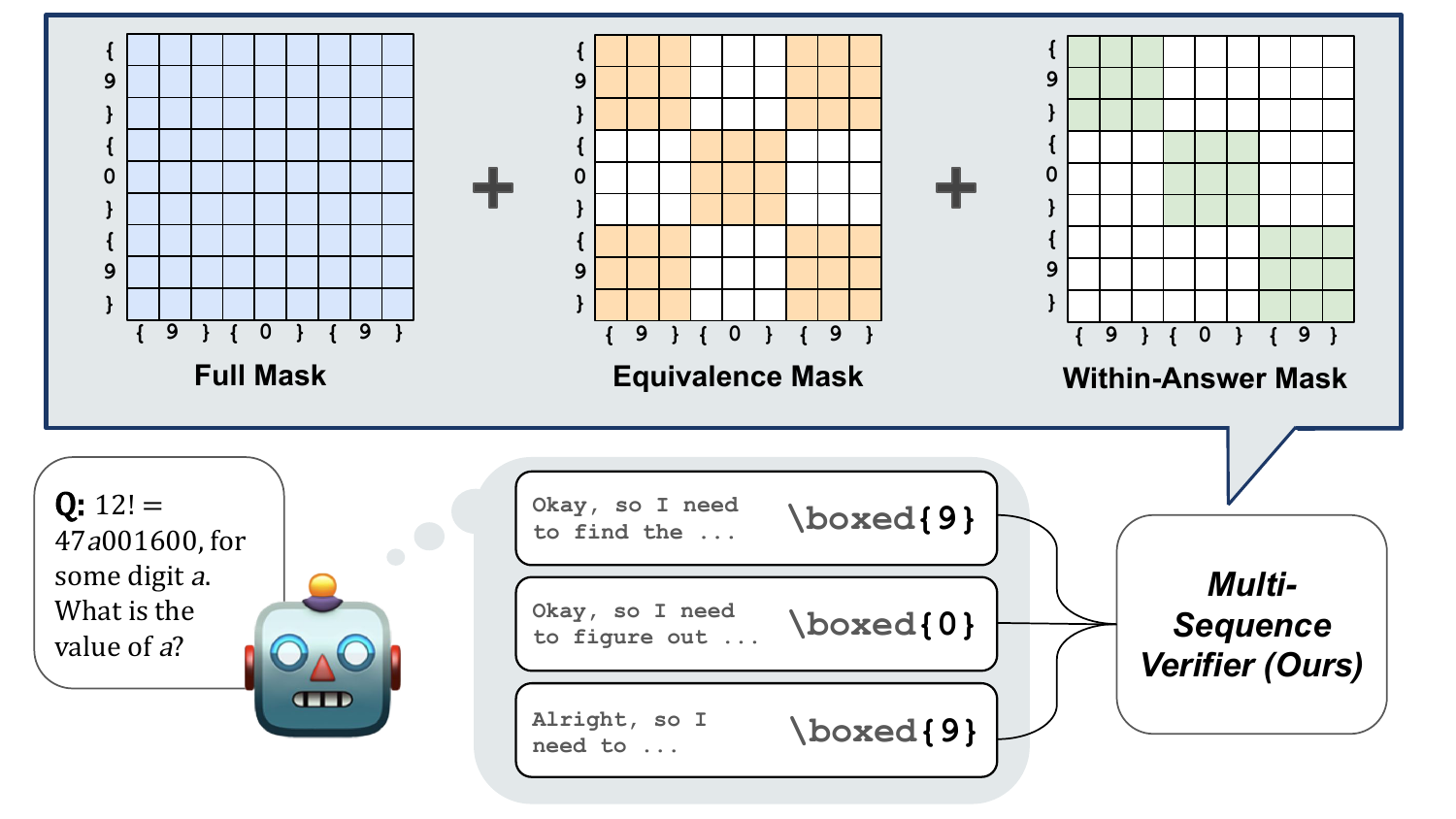}
    \caption{Illustration of our Multi-Sequence Verifier (MSV) that uses multiple attention masks in its Multi-Mask Transformer Block (MMTB) to predict the correctness of each answer. The different attention masks allow MSV to flexibly leverage information both across and within sequences.}
    \label{fig:mask_schematic}
\end{figure*}

%% file: main/experiments.tex
\section{Experiments}
\label{sec:exp}


\subsection{Experimental Setup}
\label{subsec:exp_setup}


In this section, we present empirical evidence that demonstrates the effectiveness of MSV, our novel verifier that leverages information across multiple sequences. We denote MSV trained on groups of $N$ sequences by $\text{MSV}_N$.
As for single-sequence baselines, we consider three methods:
(1)
``Probe''~\citep{zhang2025reasoning}, which is a lightweight MLP operating on last-token representation,
(2)
the open-source verifier \texttt{Qwen2.5-Math-PRM-7B} \citep[PRM;][]{zhang2025lessons}, which is a 7-billion-parameter model,
and (3)
MSV$_1$, which shares MSV$_N$'s architecture but is trained and evaluated on a single sequence at a time. MSV$_1$ serves as a controlled variant that isolates the contribution of cross-sequence information: any improvement of MSV$_N$ over MSV$_1$ can be attributed to its access to multiple candidates rather than to architectural differences.

To compare MSV against heuristic aggregation, we adopt weighted voting~\citep[WV;][]{li2022making}, which aggregates correctness probabilities by summing them within each class of symbolically equivalent answers, and normalizing the summed probabilities across classes.
In the experiments, these extensions are denoted as MSV$_1$+WV, Probe+WV, and PRM+WV.
Self-consistency~\citep{wang2022self} is a verifier-free baseline that scores each answer by its frequency among sampled solutions.

We use the \texttt{DeepSeek-R1-Distill-Qwen-1.5B} model~\citep{guo2025deepseek} as the base LM. The DeepMath-103K dataset~\citep{he2025deepmath} is used for training and validation of MSV and Probe, and evaluation is conducted on MATH~\citep{hendrycks2021measuring}, OlympiadBench~\citep[OB;][]{he2024olympiadbench}, AMC12, AIME, and Omni-MATH~\citep[OM;][]{gao2024omni}. All reported values are mean ± standard deviation over five random seeds.
We provide further details of the experimental setup in \cref{app:exp_details}, along with a full report of all experimental results in \cref{app:full_report}.

\subsection{\textbf{Terminal Answers} setting}
\label{subsec:terminal}
In this section, we conduct experiments in the \textbf{Terminal Answers} setting.
We first demonstrate, empirically, that MSV$_N$ predicts the correctness of each candidate answer more accurately than baseline methods.
We then show that, in the \emph{best-of-N} scenario, using MSV$_N$ as a verifier results in superior accuracy and calibration of the chosen answer.

\input{figure/bon/bon}

\input{table/ta_brier_bonece_small}

\paragraph{Calibration on Terminal Answers.}
\cref{fig:ta_calib} shows the Brier scores of the single-sequence baselines and MSV$_N$.
The full results with all calibration metrics can be found in \cref{tab:ta_auroc_brier,tab:ta_ece_nll}.
MSV$_N$ consistently achieves better calibration compared to Probe and MSV$_1$.
Notably, MSV$_{64}$ achieves a $\sim$50\% reduction in Brier score compared to Probe, on the AIME dataset.
Consistent with its role as a controlled variant, MSV$_1$ performs similarly to or worse than Probe, showing that the transformer block's additional capacity offers no advantage when there is no cross-sequence structure to model.

\paragraph{Best-of-N prediction.}
We apply MSV to best-of-N answer selection. As shown in \cref{fig:bon}, MSV$_N$ improves best-of-N accuracy over single-sequence baselines, with gains becoming more pronounced starting from $N=16$.
Notably, at $N=64$, MSV$_{64}$ achieves best-of-N accuracy of 58.7\% on AMC12, compared to the best baseline accuracy of 55.2\% achieved by PRM and an average baseline accuracy of 51.3\%.
We also see that self-consistency performs the worst across all datasets, and weighted voting (WV) does not provide a consistent improvement. WV degrades the performance of PRM on all datasets, and degrades Probe on the two most challenging datasets, AIME and Omni-MATH.
The failure of WV is illustrated by an example in \cref{fig:example_probs}: Probe initially assigns the highest score to the single correct solution, but WV overly suppresses it due to its low voting count of one.
These results altogether highlight the advantage of MSV over single-sequence verifiers and heuristic aggregation methods.

\paragraph{Calibrated best-of-N scores.}
Beyond obtaining the correct answer on a problem, outputting a calibrated confidence level on the chosen answer is crucial in risk-sensitive settings.
The reliability diagrams \citep{guo2017calibration} in \cref{fig:bon_calibration} and the calibration metrics reported in \cref{tab:calib_bon_only_reordered_terminal} show that MSV$_N$ is more calibrated on its own chosen answers than the single-sequence baselines are on their chosen answers. In summary, modeling cross-sequence interactions with MSV$_N$ improves the accuracy of best-of-N answers, and also delivers more reliable confidence estimates, making it favorable for downstream decision-making.


\subsection{\textbf{Streaming Answers}}
\label{sub:streaming_answer}


Next, we conduct experiments in the \textbf{Streaming Answers} setting. Unlike \textbf{Terminal Answers}, this setting involves early stopping, and accurately predicting the correctness of intermediate answers is tied to both accuracy \emph{and} efficiency.
We use the ``Wait'' token as the delimiter, and experiments with two other delimiters in \cref{app:delimiter_sensitivity} show that the trends are insensitive to the choice.


\paragraph{Calibration on Streaming Answers.}
\cref{tab:sa_standard_auroc_brier} shows that, like in the terminal answers setting, MSV$_N$ performs much better than single-sequence baselines in terms of calibration.
\cref{fig:sa_calib_time} shows the Brier score of answers averaged per token-position bins, on AIME.
We see that MSV$_{64}$ improves over baselines at every range, and especially later in the sequences.
The calibration at later token positions is more important in practice, since we typically want to stop at later positions to retain a reasonable level of accuracy.

\input{figure/sa_tradeoff_comp/sa_tradeoff_comp}

\input{figure/example_probs/group811}

\paragraph{Parallel early-stopping.}
Next, we investigate whether the superior calibration on intermediate answers transfers to improved performance in parallel early stopping.
As described in \cref{subsec:applications}, sliding $\lambda$ between 0 and 1 creates an accuracy-token tradeoff curve for each verifier, which we plot in \cref{fig:sa_tradeoff_token_position}.
On all datasets other than Omni-MATH, MSV$_{64}$ consistently represents the Pareto frontier. On MATH, we find that the maximum achievable accuracy with single-sequence baselines can be achieved by MSV$_{64}$ with less than half the token budget.

\begin{wrapfigure}[8]{r}{0.28\textwidth} 
    \vspace{-20pt}
    \centering
    \includegraphics[width=0.26\textwidth]{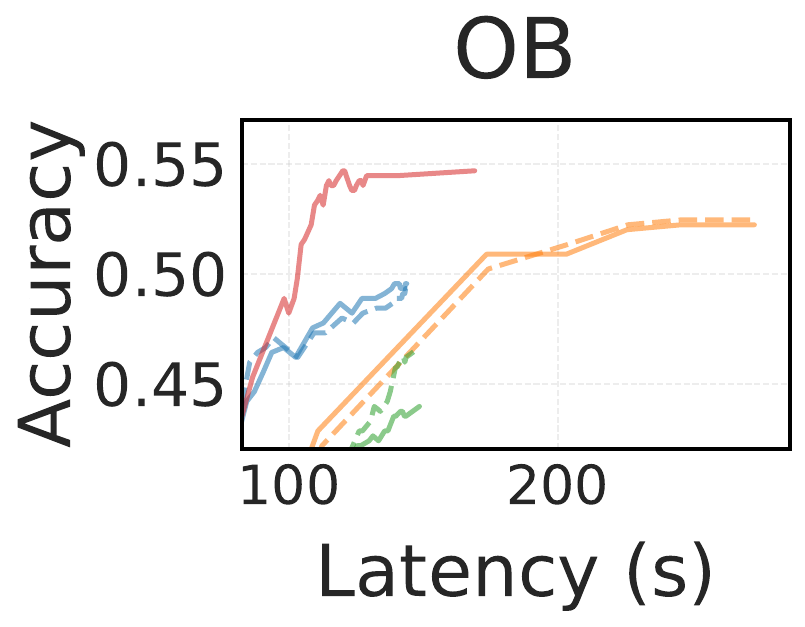}
    \caption{Accuracy-latency tradeoff with $N=64$.}
    \label{fig:math_accuracy_latency}
    \vspace{-12pt}
\end{wrapfigure}

We also provide a plot with latency as the horizontal axis, in \cref{fig:math_accuracy_latency}. PRM incurs a significant latency due to the load of a 7-billion-parameter model. As a result, while PRM seemed to provide a better tradeoff than Probe in \cref{fig:sa_tradeoff_token_position}, it is actually much worse in practice considering the tradeoff with latency. On the other hand, MSV$_{64}$ represents the Pareto frontier even in the accuracy-latency tradeoff, due to its relatively lightweight structure. We present the accuracy-latency tradeoff curves for all datasets in \cref{fig:sa_tradeoff_comp}, where we observe a similar trend.

\begin{table*}[t]
\centering
\small

\begin{minipage}[t]{0.48\textwidth}
\centering
\setlength{\tabcolsep}{2pt}
\renewcommand{\arraystretch}{1.02}
\captionof{table}{Best-of-64 accuracy on Omni-MATH under different equivalence checkers.
Full results are in \cref{tab:full_checker_robustness}.
}
\label{tab:checker_robustness}
\begin{adjustbox}{max width=\textwidth}
\begin{tabular}{lccc}
\toprule
\textbf{Checker} & \makecell{\textbf{Checker}\\\textbf{accuracy}} & \textbf{Probe+WV} & \textbf{MSV$_{64}$} \\
\midrule
SymPy (reference)       & 100.0\%           & 0.321\PM{0.002} & \textbf{0.348}\PM{0.007} \\
Embedding ($\geq 0.85$) & \phantom{0}89.4\% & 0.301\PM{0.002} & \textbf{0.344}\PM{0.008} \\
Random                  & \phantom{0}46.7\% & 0.246\PM{0.001} & \textbf{0.336}\PM{0.015} \\
\bottomrule
\end{tabular}
\end{adjustbox}
\end{minipage}
\hfill
\begin{minipage}[t]{0.48\textwidth}
\centering
\setlength{\tabcolsep}{2pt}
\renewcommand{\arraystretch}{0.95}
\captionof{table}{
Brier score and best-of-64 across different base LMs.
Full results are in \cref{tab:base_lm_calib,tab:base_lm_bo64}.
}
\label{tab:base-lm-robustness}
\begin{adjustbox}{max width=\textwidth}
\begin{tabular}{lcccc}
\toprule
& \multicolumn{2}{c}{\textbf{Brier}} 
& \multicolumn{2}{c}{\textbf{Best-of-64}} \\
\cmidrule(lr){2-3} \cmidrule(lr){4-5}
\textbf{Base LM} & \textbf{Best} & \textbf{MSV$_{64}$} 
& \textbf{Best} & \textbf{MSV$_{64}$} \\
\midrule
R1-8B     & 0.067\PM{0.010} & \textbf{0.036}\PM{0.009} & 0.649\PM{0.009} & \textbf{0.663}\PM{0.005} \\
Qwen3     & 0.077\PM{0.006} & \textbf{0.051}\PM{0.025} & 0.529\PM{0.009} & \textbf{0.536}\PM{0.011} \\
Llama 3.2 & 0.132\PM{0.005} & \textbf{0.057}\PM{0.006} & 0.338\PM{0.005} & \textbf{0.378}\PM{0.005} \\
\bottomrule
\end{tabular}
\end{adjustbox}
\end{minipage}

\end{table*}

\subsection{Robustness to Equivalence Checker Quality}
\label{sec:checker_robustness}

{
Self-consistency, weighted voting, and MSV use SymPy to group answers into equivalence classes.
We ask whether the reliance on an oracle symbolic checker is strictly necessary.
We therefore investigate the effect of using alternative checkers on best-of-N performance, while still using SymPy as the oracle for measuring correctness.
Importantly, MSV is not only evaluated but also trained using the alternative checkers so that it can learn how to best use the potentially erroneous equivalence signals. Further details on the experiments can be found in \cref{sec:full_report_checker_robustness}.
}

{
\cref{tab:checker_robustness} shows the results on Omni-MATH, with two alternative checkers---embedding cosine similarity with threshold $0.85$, and random coin flip.
MSV$_{64}$ is remarkably stable across checkers:
At 89\% checker accuracy, the drop relative to SymPy is within $0.4$ percentage points, and even with the random checker, the drop is within $1.2$ percentage points.
On the other hand, Probe+WV degrades by $2.0$ percentage points with the embedding similarity checker, and by $7.5$ percentage points with the random checker, essentially collapsing to the average accuracy of the base language model.
These results highlight the brittleness of fixed heuristic aggregation methods, and the robustness of MSV which can learn to use weak equivalence signals end-to-end.
The full results with five alternative checkers can be found in \cref{tab:full_checker_robustness}.
}

\subsection{Different Base Language Models}
\label{sec:base_lm}

{
The experiments above use \texttt{DeepSeek-R1-Distill-Qwen-1.5B} as the base language model.
To test whether the gains from MSV are specific to this setting, we repeat the Terminal Answers
experiments with three additional LMs: \texttt{DeepSeek-R1-Distill-Llama-8B}, \texttt{Qwen3-1.7B} in
thinking mode, and \texttt{Llama-3.2-1B-Instruct}. We use DeepMath-103K and AIME dataset as the training and evaluation sets, respectively, for the first two models. Since \texttt{Llama-3.2-1B-Instruct} obtains near-zero accuracy on AIME, we train and evaluate it on MATH.
}

{
\cref{tab:base-lm-robustness} summarizes the results. It compares MSV$_{64}$ against the best single-sequence baseline (``Best'') in each scenario. The full results are available in \cref{app:base_lm_generalization}.
Across all three base LMs, MSV$_{64}$ reduces the Brier score to around half that of the best baselines, and consistently improves best-of-64 accuracy. These results indicate that the superiority of MSV over single-sequence verifiers generalizes to different model sizes and families, and to non-reasoning models.
}

%% file: figure/bon/bon.tex
\begin{figure*}[t]
    \centering
    \includegraphics[width=1.0\textwidth]{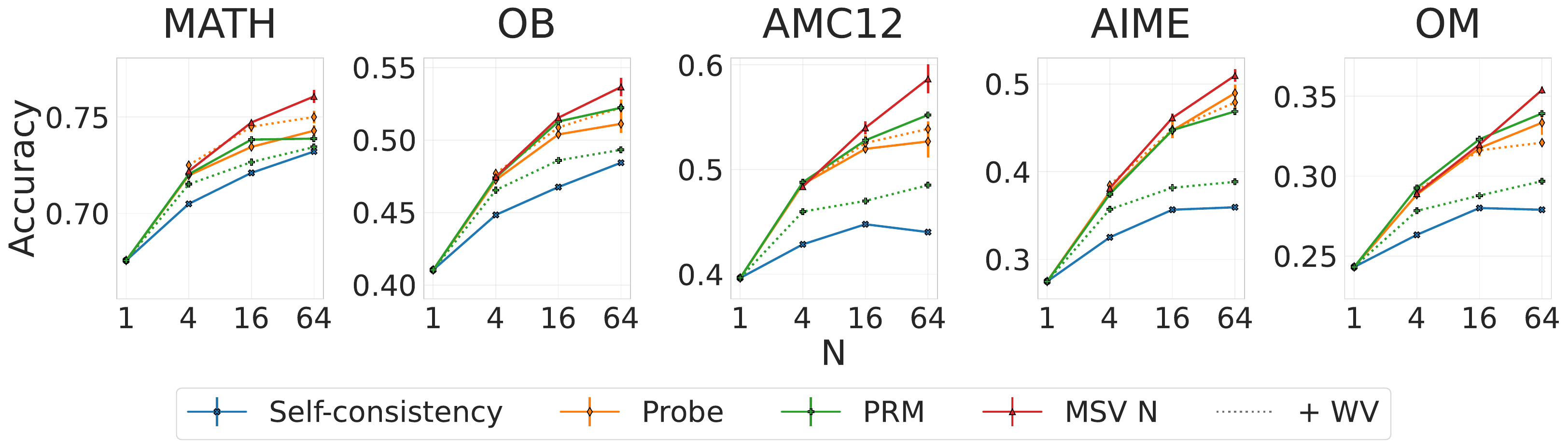}
    \caption{Best-of-N accuracy vs. $N$, in \textbf{Terminal Answers} setting. Full results are in \cref{tab:acc_bon_only_reordered_terminal}.}
    \label{fig:bon}
\end{figure*}

%% file: table/ta_brier_bonece_small.tex
\begin{figure}[t]
\centering
\begin{minipage}[c]{0.53\textwidth}
\centering
\includegraphics[width=\linewidth]{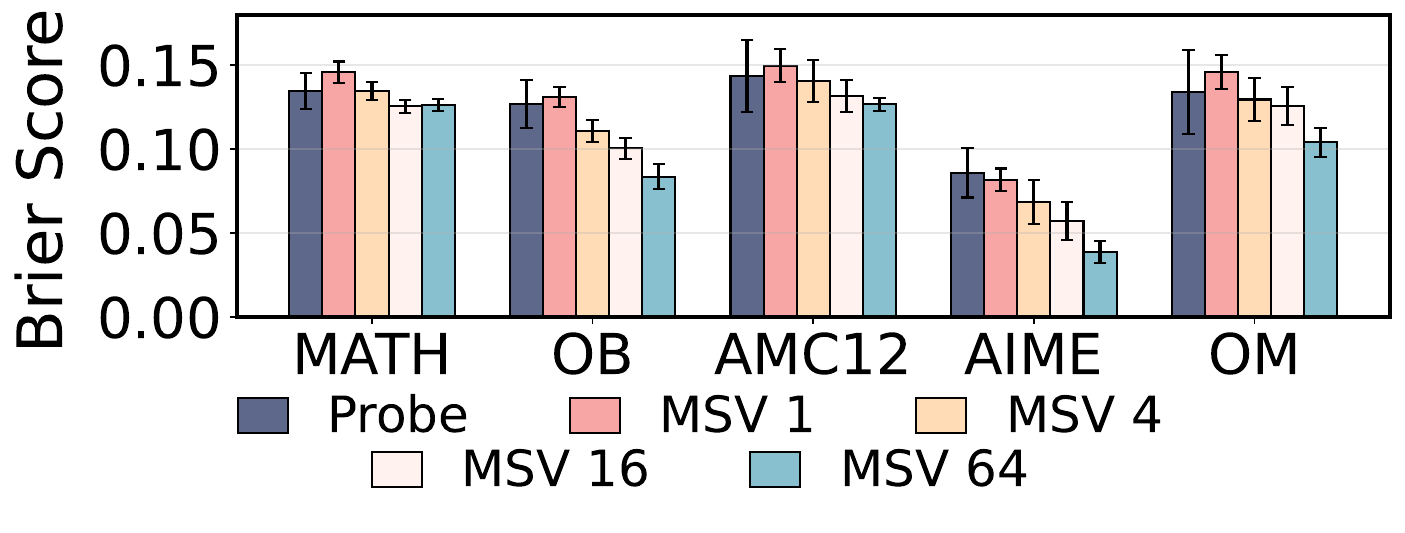}
\captionof{figure}{Brier scores ($\downarrow$) in the \textbf{Terminal Answers} setting. Full results are in \cref{tab:ta_auroc_brier}.}
\label{fig:ta_calib}
\end{minipage}\hfill
\begin{minipage}[c]{0.42\textwidth}
\centering
\includegraphics[width=\linewidth]{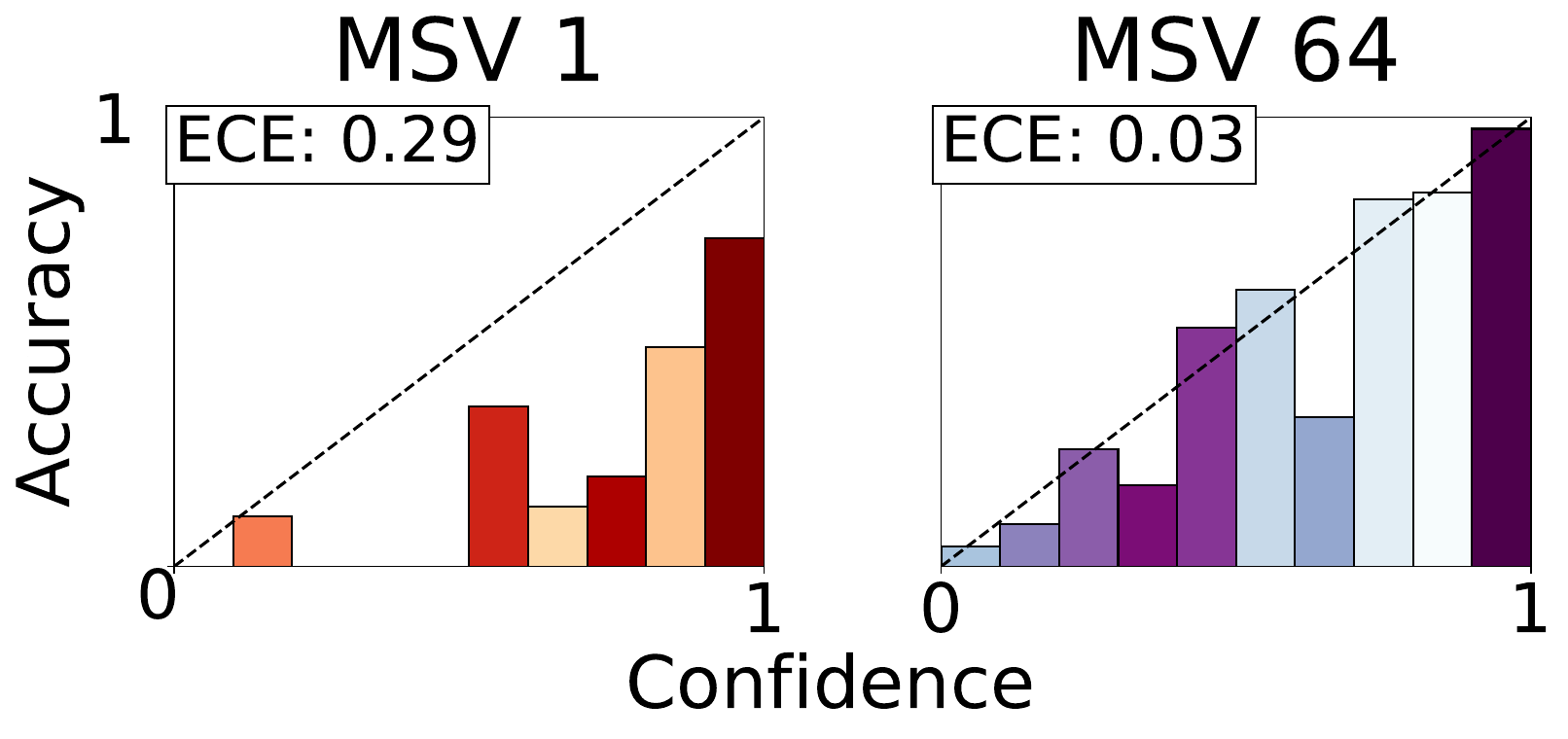}
\captionof{figure}{Reliability diagrams of verifier confidence on best-of-64 answers (AIME).}
\label{fig:bon_calibration}
\end{minipage}
\end{figure}




%% file: figure/sa_tradeoff_comp/sa_tradeoff_comp.tex
\begin{figure}[t]
    \centering
    \includegraphics[width=0.98\linewidth]{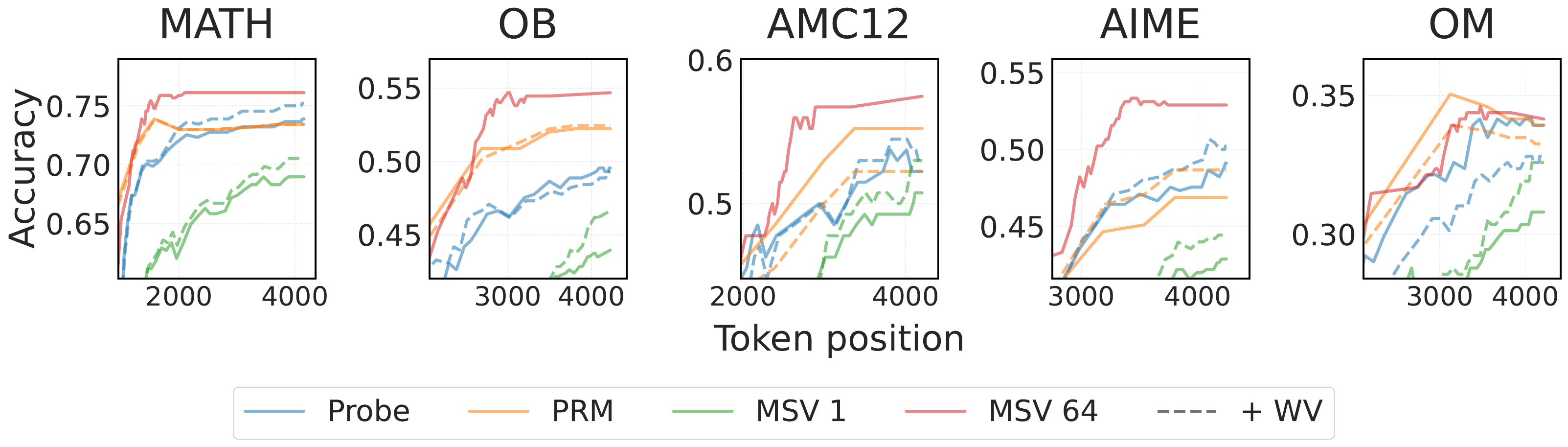}
    \caption{Accuracy-token tradeoff curves in \textbf{Streaming Answers}, with $N=64$.
    Curves that lie higher represent superior tradeoff. Accuracy-latency tradeoff curves are in \cref{fig:sa_tradeoff_comp}.}
    \label{fig:sa_tradeoff_token_position}
\end{figure}



%% file: figure/example_probs/group811.tex
\begin{figure*}[t]
    \centering
    \begin{minipage}[t]{0.7\textwidth}
        \centering
        \includegraphics[width=\linewidth]{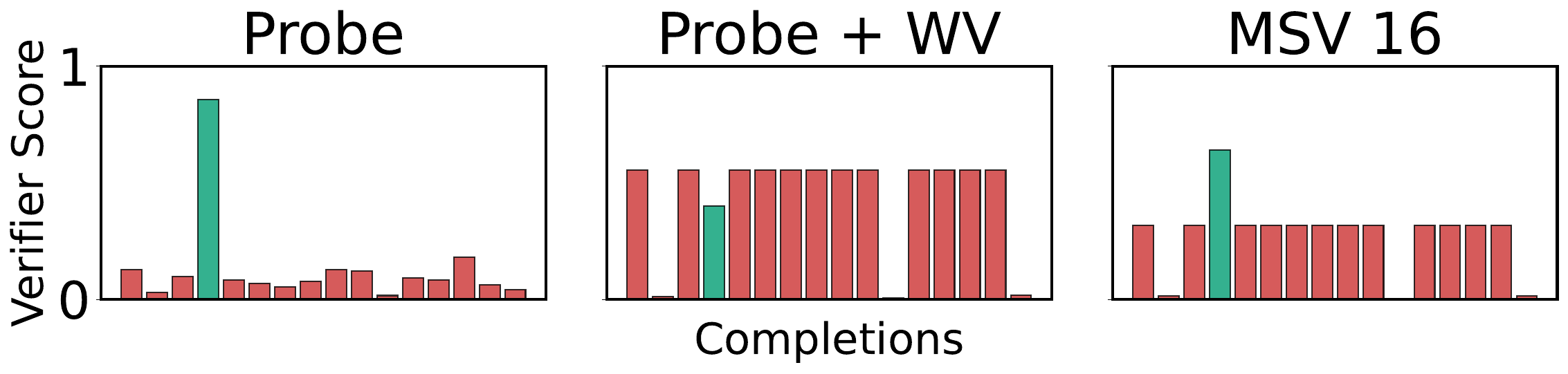}
        \caption{Probability estimates from Probe, Probe+WV, and MSV$_{16}$ for a single problem in Omni-MATH, with 16 completions. Green represents the correct solution.}
        \label{fig:example_probs}
    \end{minipage}
    \hfill
    \begin{minipage}[t]{0.26\textwidth}
        \centering
        \includegraphics[width=0.98\linewidth]{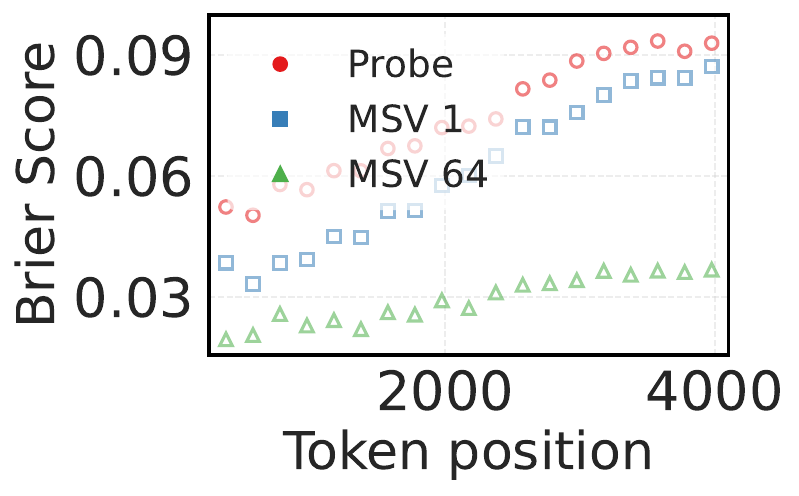}
        \caption{Brier score per token position, on AIME, Streaming Answers.
        }
        \label{fig:sa_calib_time}
    \end{minipage}
\end{figure*}

%% file: main/conclusion.tex
\section{Conclusion}
\label{sec:conclusion}


In this work, we addressed two critical bottlenecks in parallel test-time scaling: the accurate selection of correct solutions, and the high inference latency from generating multiple candidates. We argued that both challenges are fundamentally linked to verifier calibration and that existing verifiers are limited by scoring candidate solutions in isolation. To overcome this, we introduced the Multi-Sequence Verifier (MSV), a novel architecture designed to jointly process an entire set of candidate solutions and model their interactions. Our extensive experiments demonstrated that MSV achieves superior calibration, significantly outperforming strong baselines that score sequences independently. This improved calibration directly translated to substantial downstream benefits, such as accurate best-of-N predictions and a calibrated confidence score associated with each prediction.

Furthermore, we introduced a novel parallel early-stopping framework for efficient inference that contrasts with prior sequential approaches. In this framework, a streaming variant of MSV achieved the same peak accuracy as the best baseline verifier on MATH with around half the latency. These findings underscore the importance of cross-sequence information and end-to-end training, and establish a new, effective approach to building and using verifiers for parallel test-time scaling.

\paragraph{Limitations.}
We discuss several limitations of our work.
MSV's advantage over single-sequence baselines is not uniform across all values of $N$, with clear and consistent gains emerging from $N=16$ onward.
Our experiments are also narrow in scope, focusing on mathematics datasets, albeit on a wide range of difficulties.
The computational complexity of MSV$_N$ is quadratic in $N$, and we propose an effective workaround in \cref{sec:scaling} with empirical evidence.
MSV also requires access to internal hidden states of a language model.
Finally, MSV relies on an equivalence checker, and high-quality equivalence checkers might not always be available. We show in \cref{sec:checker_robustness} that MSV is still extremely robust to approximate equivalence checkers.

%% file: appendix/experimental_details.tex
\section{Experimental Details}
\label{app:exp_details}

\subsection{Details on Baseline methods}
\label{app:baselines}

\subsubsection{Single-Sequence Baselines}
To show that the interaction between sequences really does benefit the classification performance of \gls{msv}, we compare \gls{msv} trained on $N>1$ against \gls{msv} trained on $N=1$, which serves as a controlled baseline. We abbreviate \gls{msv} trained on $N$ sequences simply as \gls{msv}$_N$. 
As another baseline, we follow \citet{zhang2025reasoning} and train a 2-layer MLP ``Probe'' on the last token representations $h\nth_{k,L\nth_k}$ to predict the correctness of answers $a\nth_k$. We find that the Probe serves as a strong baseline that often matches or even exceeds the performance of \gls{msv}$_1$.

We also consider the open-source verifier \texttt{Qwen2.5-Math-PRM-7B} \citep[PRM;][]{zhang2025lessons}, which is a 7-billion-parameter process reward model trained to score intermediate reasoning steps. Intermediate reasoning traces with special tokens \texttt{<extra\_0>} are fed to the PRM, which outputs correctness scores for the traces; we use the softmax probability of the positive class as the correctness score. Specifically, in both the Terminal Answers and Streaming Answers setting, we feed to the PRM the currently decoded reasoning trace appended with the terminal/intermediate answer to evaluate.

We focus on non-generative, white-box methods in our paper. Generative verifiers~\citep{zhang2024generative} and joint-selection approaches such as USC \citep{chen2023universal}, PairJudge RM \citep{liu2025pairjudge}, and GenSelect \citep{toshniwal2025genselect} share the intuition of evaluating candidates in context of one another, but their inference costs are orders of magnitude higher: GenSelect requires approximately 130 seconds to judge a group of 16 solutions versus 0.006 seconds for MSV$_{16}$. This makes them inapplicable in our \textbf{Streaming Answers} setting, where a judgment must be issued at every intermediate answer. Multi-agent debate \citep{du2023improving} faces a similar cost issue. The training scheme of these methods also differs substantially, often involving reinforcement learning or prompting rather than supervised learning on hidden states.

There are also training-free methods for calibration, such as using the base LLM's token probabilities on the answer tokens~\citep{yang2025dynamic}. Specifically, we use the geometric mean of probabilities over all tokens in an answer, to estimate the answer's correctness.
\[
s\nth_k = \left(\prod_{i=1}^{L\nth_k} p_{\textrm{LLM}}\left(a\nth_{k,i} \mid \textrm{tokens}_{<a\nth_{k,i}}\right)\right)^{1/L\nth_k}.
\]

\subsubsection{Aggregation Baselines}
\label{sec:aggregation_baselines}
We can further complement the single-sequence baselines with aggregation methods that combine predictions across multiple sequences through simple heuristics. Although prior work has not investigated explicit aggregation of single-sequence baselines for improving the classification of individual candidates, we introduce a baseline, \emph{Weighted Voting} (WV), inspired by the weighted best-of-N approach~\citep{li2022making}, as follows:
for each equivalence class of symbolically identical answers, we aggregate its answers' correctness probabilities by summing them, and normalize the probabilities across equivalence classes. Performing best-of-N with the WV probabilities is equivalent to weighted best-of-N, which picks the equivalence class with the biggest aggregate probability. In the streaming setting, we can also perform weighted voting for candidate $a\nth_k$ by aggregating and normalizing over the symbolically equivalent ones that come before time $\tau\nth_k$.
However, when done naively, this can disrupt the predictions at the early token positions, due to a lack of other existing candidates. For instance, weighted voting will always assign a probability of one to the first output candidate, since there are no other candidates to normalize over. Therefore, we set a threshold $R\in \mathbb Z$ such that weighted voting is performed only when the number of candidates exceeds $R$ at time $\tau\nth_k$. We find that $R=16$ yields good overall performance and use this setting throughout our experiments with WV in the \textbf{Streaming Answers} setting.

We also evaluate the training-free baseline \emph{self-consistency} \citep{wang2022self} in the Terminal Answers setting that only uses the vote count (number of symbolically identical answers) to score each answer. Specifically, we score each answer $a\nth$ with
\[
s\nth = \sum_{m=1}^N \mathbf1 [ a\nth \sim a\smallth{m} ]  / N.
\]


\subsection{Details on Datasets and Models}
\label{app:datasets_and_models}
The pre-trained model and datasets used in our experiments are summarized below.


\textbf{Models.}
\begin{itemize}[leftmargin=0.6cm,nosep]
  \item \textbf{\texttt{DeepSeek-R1-Distill-Qwen-1.5B}}:
    \href{https://huggingface.co/deepseek-ai/DeepSeek-R1-Distill-Qwen-1.5B}{https://huggingface.co/deepseek-ai/DeepSeek-R1-Distill-Qwen-1.5B},
    licensed under MIT License.\footnote{\href{https://huggingface.co/deepseek-ai/DeepSeek-R1-Distill-Qwen-1.5B/blob/main/LICENSE}{https://huggingface.co/deepseek-ai/DeepSeek-R1-Distill-Qwen-1.5B/blob/main/LICENSE}}

  \item \textbf{\texttt{DeepSeek-R1-Distill-Llama-8B}}:
    \href{https://huggingface.co/deepseek-ai/DeepSeek-R1-Distill-Llama-8B}{https://huggingface.co/deepseek-ai/DeepSeek-R1-Distill-Llama-8B},
    licensed under MIT License.\footnote{\href{https://huggingface.co/deepseek-ai/DeepSeek-R1-Distill-Llama-8B/blob/main/LICENSE}{https://huggingface.co/deepseek-ai/DeepSeek-R1-Distill-Llama-8B/blob/main/LICENSE}}

  \item \textbf{\texttt{Qwen3-1.7B}}:
    \href{https://huggingface.co/Qwen/Qwen3-1.7B}{https://huggingface.co/Qwen/Qwen3-1.7B},
    licensed under Apache License 2.0.\footnote{\href{https://huggingface.co/Qwen/Qwen3-1.7B/blob/main/LICENSE}{https://huggingface.co/Qwen/Qwen3-1.7B/blob/main/LICENSE}}
    We use \texttt{Qwen3-1.7B} in thinking mode.

  \item \textbf{\texttt{Llama-3.2-1B-Instruct}}:
    \href{https://huggingface.co/meta-llama/Llama-3.2-1B-Instruct}{https://huggingface.co/meta-llama/Llama-3.2-1B-Instruct},
    licensed under the Llama 3.2 Community License.\footnote{\href{https://huggingface.co/meta-llama/Llama-3.2-1B-Instruct/blob/main/LICENSE.txt}{https://huggingface.co/meta-llama/Llama-3.2-1B-Instruct/blob/main/LICENSE.txt}}

  \item \textbf{\texttt{Qwen2.5-Math-PRM-7B}}:
    \href{https://huggingface.co/Qwen/Qwen2.5-Math-PRM-7B}{https://huggingface.co/Qwen/Qwen2.5-Math-PRM-7B},
    licensed under the Qwen License Agreement.\footnote{\href{https://huggingface.co/Qwen/Qwen2.5-Math-PRM-7B/blob/main/LICENSE}{https://huggingface.co/Qwen/Qwen2.5-Math-PRM-7B/blob/main/LICENSE}}
\end{itemize}

\textbf{Datasets.}
\begin{itemize}[leftmargin=0.6cm,nosep]

\item \textbf{\texttt{MATH-500}}:
\href{https://huggingface.co/datasets/HuggingFaceH4/MATH-500}{https://huggingface.co/datasets/HuggingFaceH4/MATH-500},
    licensed under MIT License.\footnote{\href{https://github.com/openai/prm800k/blob/main/LICENSE}{https://github.com/openai/prm800k/blob/main/LICENSE}}

\item \textbf{\texttt{Olympiad Bench}}:
\href{https://huggingface.co/datasets/Hothan/OlympiadBench}{https://huggingface.co/datasets/Hothan/OlympiadBench},
    licensed under MIT License.\footnote{\href{https://github.com/OpenBMB/OlympiadBench/blob/main/LICENSE}{https://github.com/OpenBMB/OlympiadBench/blob/main/LICENSE}}

\item \textbf{\texttt{AMC12}}:
\href{https://huggingface.co/datasets/rulins/amc12_22-24}{https://huggingface.co/datasets/rulins/amc12\_22-24},
    not openly licensed; usage is subject to MAA AMC policies.\footnote{\href{https://maa.org/student-programs/amc/maa-american-mathematics-competitions-policies/}{https://maa.org/student-programs/amc/maa-american-mathematics-competitions-policies/}}

\item \textbf{\texttt{AIME}}:
\href{https://www.kaggle.com/datasets/hemishveeraboina/aime-problem-set-1983-2024}{https://www.kaggle.com/datasets/hemishveeraboina/aime-problem-set-1983-2024},\\
licensed under CC0 License.
\footnote{\href{https://www.kaggle.com/datasets/hemishveeraboina/aime-problem-set-1983-2024}{https://www.kaggle.com/datasets/hemishveeraboina/aime-problem-set-1983-2024}}

\item \textbf{\texttt{Omni-MATH}}:
\href{https://huggingface.co/datasets/KbsdJames/Omni-MATH}{https://huggingface.co/datasets/KbsdJames/Omni-MATH},\\
dataset files licensed under Apache-2.0 License.\footnote{\href{https://huggingface.co/datasets/KbsdJames/Omni-MATH}{https://huggingface.co/datasets/KbsdJames/Omni-MATH}}
    
\item \textbf{\texttt{DeepMath-103K}}:
\href{https://huggingface.co/datasets/zwhe99/DeepMath-103K}{https://huggingface.co/datasets/zwhe99/DeepMath-103K},
    licensed under MIT License.\footnote{\href{https://github.com/zwhe99/DeepMath/blob/main/LICENSE}{https://github.com/zwhe99/DeepMath/blob/main/LICENSE}}
 
\end{itemize}

We use the DeepSeek-R1-Distill-Qwen-1.5B model~\citep{guo2025deepseek} as our base LLM that generates the reasoning traces and answers to reasoning problems.
All training data for our verifier models are generated with the base LLM on problems in the DeepMath-103K dataset, which underwent careful decontamination of public reasoning evaluation datasets, including all five of our evaluation datasets~\citep{he2025deepmath}. This reduces the possibility that the verifier is memorizing the training data to correctly classify solutions generated on our evaluation sets.
We evaluate our model on the MATH dataset~\citep{hendrycks2021measuring}, the AMC12 problems between 2022 and 2024 that are not geometry problems, the AIME problems between 1983 and 2024, OlympiadBench~\citep{he2024olympiadbench}, and Omni-MATH~\citep{gao2024omni}.
We use the full 1983--2024 AIME dataset rather than AIME 2024/2025, because the latter contains only 60 problems, which is too small for statistically reliable evaluation. The DeepMath-103K training set was decontaminated to exclude all problems from the full AIME dataset~\citep{he2025deepmath}, so there is no data leakage risk for verifier training.

We use 224 and 64 randomly chosen problems from DeepMath-103K as our training and validation sets, respectively. Each evaluation dataset is randomly subsampled to 448 problems (except for AMC12, which has 134 problems). We generate 64 responses from the base LLM on every problem. This creates training and validation sets with 14K and 4K sequences, respectively, and evaluation datasets with 28K sequences.
We provide justification for the training set size in \cref{app:training_set_size}.

\subsection{Details on metrics}
\label{app:metrics}
We evaluate verifier predictions on four standard probabilistic metrics. Let $\{(p_i, y_i)\}_{i=1}^m$ denote the dataset of $m$ predicted probabilities and their corresponding binary correctness labels.

\paragraph{Area Under the Receiver Operating Characteristic Curve (AUROC).}
The Area Under the Receiver Operating Characteristic Curve (AUROC) measures the probability that a randomly chosen positive example receives a higher score than a randomly chosen negative example:
\[
\mathrm{AUROC} = \Pr\big[p^+ > p^-\big] + \frac{1}{2}\Pr\big[p^+ = p^-\big],
\]

where $p^+$ and $p^-$ are the predicted probabilities for a random positive and negative sequence, respectively.

\paragraph{Brier score.}
The Brier score \citep{brier} is the mean squared error between predicted probabilities and binary outcomes:
\[
\mathrm{Brier} = \frac{1}{m}\sum_{i=1}^m (p_i - y_i)^2.
\]
Lower values indicate better calibrated and sharper probabilities.

\paragraph{Negative Log-Likelihood (NLL).}
The negative log-likelihood, or log loss, evaluates the quality of probabilistic predictions:
\[
\mathrm{NLL} = -\frac{1}{m}\sum_{i=1}^m \Big( y_i \log p_i + (1-y_i)\log (1-p_i) \Big).
\]

\paragraph{Expected Calibration Error (ECE).}
The expected calibration error \citep{naeini2015obtaining, guo2017calibration} measures the discrepancy between a model's predicted confidences and the actual accuracies. To compute ECE, predictions are grouped into $J$ bins based on their confidence scores. The ECE is the weighted average of the absolute difference between the mean confidence and the mean accuracy within each bin:
\[
\mathrm{ECE} = \sum_{j=1}^{J} \frac{|B_j|}{m} \left| \mathrm{acc}(B_j) - \mathrm{conf}(B_j) \right|,
\]
where $B_j$ is the set of indices of predictions whose confidence $p_i$ falls into the $j$-th bin, $\mathrm{acc}(B_j) = \frac{1}{|B_j|}\sum_{i \in B_j} y_i$ is the accuracy of that bin, and $\mathrm{conf}(B_j) = \frac{1}{|B_j|}\sum_{i \in B_j} p_i$ is the average confidence of that bin. We use $J=10$ in all our reports of ECE.

\paragraph{Best-of-N Accuracy.}
Best-of-N Accuracy is the primary metric for evaluating the effectiveness of the verifier-guided selection process in the Terminal Answers setting. For each problem, the system generates $N$ candidate answers. A verifier then provides a score for each candidate, and the one with the highest score $p^\dagger$, denoted $a^\dagger$, is selected as the final output. The accuracy is the fraction of problems in an evaluation set $Q$ for which this selected answer is equivalent to the ground truth answer $a^*$. Formally, it is expressed as:
\[
\text{Accuracy} = \frac{1}{|Q|} \sum_{q \in Q} y^\dagger_q, \, \text{ where } y^\dagger_q=\mathbf{1}[a_q^\dagger \sim a_q^*]
\]
and $a_q^\dagger$ is the verifier-selected answer for problem $q$, and $a_q^*$ is the ground truth. A higher accuracy indicates a more effective verifier that is better at identifying the correct solution from the pool of candidates, thus serving as a direct measure of downstream task performance.

\paragraph{Best-of-N ECE and Brier score.}
Best-of-N ECE and Brier scores are the ECE and Brier scores computed over the set $\{ (p^\dagger_q, y^\dagger_q) \}$ of best-of-N confidence and correctness pairs.


\subsection{Details on Experiments}
We adopt the setup and terminology of Terminal/Streaming answers from \cref{sec:method}, and follow the experimental protocol of \cref{subsec:exp_setup}. Below we provide additional details necessary for reproducibility.

\textbf{{Optimizer and schedule.}} 
All models are trained with AdamW, combined with the Hugging Face default \texttt{constant-with-warmup} scheduler (default warmup ratio). Gradient clipping is applied with \texttt{max\_grad\_norm} set to $1.0$.

\textbf{{Learning rate selection.}}
We sweep learning rates in $\{1\text{e}{-5},\,5\text{e}{-5},\,1\text{e}{-4},\,5\text{e}{-4},\,1\text{e}{-3},\,5\text{e}{-3}\}$ and choose the best value by Brier score cross-validation on the validation split.

\textbf{{Final learning rates.}} 
The probe MLP is trained with $\text{lr}=1\text{e}{-3}$. 
All other modules use $\text{lr}=5\text{e}{-5}$, except the MMTB mixture weights $w$ ($\text{lr}=1\text{e}{-1}$) and the per-sequence embeddings $e^{(n)}$ ($\text{lr}=1\text{e}{-3}$).  

\textbf{{Model capacity.}} 
The MSV verifier consists of a single Multi-Mask Transformer Block (one transformer layer) whose width and attention heads follow the base model. The probe is implemented as a 2-layer MLP with hidden size $1024$.  

\textbf{{Batching and epochs.}}
We use a global batch size of $64$ with no gradient accumulation. Training runs for $1$ epoch on Terminal data and $2$ epochs on Streaming data.

\textbf{{Decoding and prompting.}}
We follow the elicitation protocol in the main paper, using the boxed \texttt{Final Answer} format and the ``\texttt{Wait}'' delimiter for Streaming. Decoding uses temperature $=1.0$ without further heuristics. Each output is allowed up to $4096$ tokens, with the elicited final answer truncated to $40$ tokens. Chain-of-thought generation uses PyTorch’s scaled dot-product attention (SDPA) backend and KV caching via HuggingFace's \texttt{generate} function.  

\textbf{{Hardware.}}
All experiments fit within a single RTX A6000 GPU (48\,GB memory) per model.  

\begin{table}[t]
\centering
\small
\setlength{\tabcolsep}{6pt}
\begin{tabular}{@{}ll@{}}
\toprule
\textbf{Category} & \textbf{Setting (ours)} \\
\midrule
Optimizer & AdamW \\
LR schedule & \texttt{constant-with-warmup} (HF defaults) \\
LR sweep & $\{1\text{e}{-5},\,5\text{e}{-5},\,1\text{e}{-4},\,5\text{e}{-4},\,1\text{e}{-3},\,5\text{e}{-3}\}$ (CV on validation) \\
Final LRs & Probe: $1\text{e}{-3}$; Others: $5\text{e}{-5}$; $w$: $1\text{e}{-1}$; $e^{(n)}$: $1\text{e}{-3}$ \\
Probe arch. & 2-layer MLP, hidden size 1024 \\
MSV depth & 1 Multi-Mask Transformer Block \\
Batch & 64 \\
Gradient accumulation steps & 1 \\
Epochs & Terminal: 1; Streaming: 2 (4 only during LR sweep) \\
Gradient clipping & \texttt{max\_grad\_norm} $=1.0$ \\
Gradient checkpointing & Not used \\
Max sequence length & 4096 \\
Temperature& 1.0\\
Answer length & $\leq 40$ tokens (after elicitation) \\
\bottomrule
\end{tabular}
\caption{Summary of implementation and hyperparameters.}
\end{table}



%% file: appendix/additional_exp.tex
\section{Additional Experiments}
\label{app:additional_exp}

\begin{table*}[t]
\caption{Average latency breakdown and end-to-end latency, in seconds.}
\label{tab:streaming_latency_all}
\centering
\setlength{\tabcolsep}{3pt}
\renewcommand{\arraystretch}{0.9}

\begin{subtable}[t]{0.15\textwidth}
\centering
\caption{CoT gen.}
\label{tab:streaming_cot_latency}
\resizebox{\linewidth}{!}{%
\begin{tabular}{lc}
\toprule
$N$ & \textbf{Latency} \\
\midrule
4  & 52.8  \\
16 & 62.4  \\
64 & 131.6 \\
\bottomrule
\end{tabular}
}
\end{subtable}
\hfill
\begin{subtable}[t]{0.15\textwidth}
\centering
\caption{Answer gen.}
\label{tab:streaming_answer_latency}
\resizebox{\linewidth}{!}{%
\begin{tabular}{lc}
\toprule
$N$ & \textbf{Latency} \\
\midrule
4  & 1.5 \\
16 & 1.8 \\
64 & 3.8 \\
\bottomrule
\end{tabular}
}
\end{subtable}
\hfill
\begin{subtable}[t]{0.27\textwidth}
\centering
\caption{Verifier inference}
\label{tab:streaming_verifier_latency}
\resizebox{\linewidth}{!}{%
\begin{tabular}{lccc}
\toprule
$N$ & \textbf{Probe} & \textbf{MSV$_N$} & \textbf{PRM} \\
\midrule
4  & 0.0008 & 0.4  & 7.4   \\
16 & 0.003  & 2.3  & 29.6  \\
64 & 0.01   & 23.2 & 118.5 \\
\bottomrule
\end{tabular}
}
\end{subtable}
\hfill
\begin{subtable}[t]{0.27\textwidth}
\centering
\caption{End-to-end latency}
\label{tab:streaming_e2e_latency}
\resizebox{\linewidth}{!}{%
\begin{tabular}{lccc}
\toprule
$N$ & \textbf{Probe} & \textbf{MSV$_N$} & \textbf{PRM} \\
\midrule
4  & 54.6  & 55.0  & 62.0  \\
16 & 64.5  & 66.8  & 94.1  \\
64 & 135.7 & 158.9 & 254.2 \\
\bottomrule
\end{tabular}
}
\end{subtable}

\end{table*}

\subsection{{Latency Analysis in the Streaming Answers Setting}}
\label{app:latency_analysis}

{
To substantiate our claim that \gls{msv} improves the accuracy-latency tradeoff compared to baselines, we provide detailed measurements of latency on the AIME dataset in the Streaming Answers setting.
Specifically, we decompose the end-to-end latency when we never early stop, i.e., set the threshold to be $\lambda=1$.
We break down the latency into three components: chain-of-thought generation, intermediate answer generation, and verifier inference.
}

{
\paragraph{Wall-Clock Time Analysis.}
We measure the actual wall-clock time for each of the three components using batched computation on a single A6000 GPU. Tables~\ref{tab:streaming_cot_latency}, \ref{tab:streaming_answer_latency}, and \ref{tab:streaming_verifier_latency} report the average latency for chain-of-thought generation, intermediate answer generation, and verifier inference, respectively. Latency measurements assume the online KV-cached implementation of MSV described in \cref{app:online_inference}.
}

\paragraph{End-to-End Latency.}
\cref{tab:streaming_e2e_latency} reports the total end-to-end latency by summing the three components plus the time taken for symbolic equivalence checks.
The table might give the impression that MSV$_N$ has higher latency than Probe, but this is because we reported the latency at $\lambda=1$, where MSV$_N$ achieves greater accuracy than Probe.
Looking at \cref{fig:sa_tradeoff_comp}, we find that MSV$_N$ achieves the same accuracy as Probe with significantly lower wall-clock latency, showing that the cost of verifier inference is more than compensated by the improved tradeoff.
On the other hand, PRM incurs substantially greater overhead, as it requires a full forward pass through a 7-billion-parameter model for each intermediate answer. This is far more expensive than MSV, which is a single transformer block on top of the existing LLM hidden states.

{
\paragraph{Training latency.}
In the Terminal Answers setting, training takes around 50 minutes regardless of the verifier. In the Streaming Answers setting, training of Probe, MSV$_1$, MSV$_4$, MSV$_{16}$, and MSV$_{64}$ takes around 105, 105, 105, 110, and 145 minutes, respectively. We performed cross-validation on the number of epochs, meaning that running more epochs on Probe, MSV$_{1}$, or MSV$_{4}$ doesn’t improve their performance.
}




\begin{table}[t]
\caption{Mask ablation results on AIME with MSV$_{16}$ in the Streaming Answers setting.}
\label{tab:mask_ablation}
\centering
\setlength{\tabcolsep}{6pt}
\renewcommand{\arraystretch}{1.0}
{
\begin{tabular}{lccccc}
\toprule
\textbf{Metric} & \textbf{MSV$_{16}$} & \textbf{w/o full} & \textbf{w/o equiv.} & \textbf{w/o within-seq} & \textbf{w/o within-ans} \\
\midrule
Brier ↓ & \textbf{0.038} & 0.040 & 0.045 & 0.048 & 0.050 \\
AUROC ↑ & \textbf{0.953} & \textbf{0.953} & 0.922 & 0.945 & 0.949 \\
NLL ↓   & \textbf{0.150} & 0.157 & 0.236 & 0.192 & 0.169 \\
\bottomrule
\end{tabular}
}
\end{table}

\subsection{{Ablation Study of Attention Masks}}
\label{sec:mask_ablation}

{
To assess the contribution of each attention mask in \gls{msv}, we conduct an ablation study by removing each mask individually and evaluating the resulting model's calibration performance. \cref{tab:mask_ablation} reports the Brier score, AUROC, and negative log-likelihood (NLL) of the Streaming MSV$_{16}$ model on AIME, with each mask ablated in turn.
}

{
The results show that all masks contribute to the overall performance, but the equivalence and within-sequence masks are especially important. Removing the equivalence mask increases the Brier score by 18.4\% and substantially degrades AUROC and NLL, indicating that cross-sequence attention to equivalent answers is critical for effective verification. Removing the within-sequence mask also leads to noticeable degradation across all metrics. In contrast, removing the full mask has the smallest effect, suggesting that the more specialized masks capture the most relevant information for verification.
}

{
\paragraph{Motivation for Multiple Masks.}
Our decision to use multiple specialized masks was informed by preliminary experiments in which we allowed the verifier to attend to all tokens in the sequence, including chain-of-thought tokens. We initially hypothesized that, since attending to more tokens strictly increases the available information, specialized masks would be unnecessary.
}

{
Contrary to this expectation, we observed that restricting attention to only the answer tokens led to substantially better generalization. This indicated that transformers can be distracted by less critical information, and that explicitly guiding attention toward the most informative parts of the sequence improves performance. Building on this insight, we tested more specialized masks that attend only to semantically related subsets of answers (e.g., equivalent answers within and across sequences). The ablation results confirm that appropriately combining these specialized masks yields the strongest overall performance, as each mask captures complementary aspects of the multi-sequence structure.
}

\subsection{Ablation: Logit Averaging in Streaming Answers}
\label{sec:logit_avg_ablation}

\input{table/main_ablation}

In \cref{subsec:msv}, we presented several design choices of MSV without extensive justification. The most notable is the absence of logit averaging in the prediction stage of the {Streaming Answers} setting. We therefore design a version of MSV that uses logit averaging during training and inference. Since the predictions have to respect causality in the {Streaming Answers} setting, we average the logits over all symbolically equivalent answers that come before in time.
We train the ablated model architecture with $N=16$, and evaluate on the AIME dataset. We report the standard calibration metrics in \cref{tab:main_ablation}. One can see that logit averaging actually hurts the performance of MSV$_{16}$, potentially due to the low quality logits in the early parts of the sequences that become aggregated even in later parts of the sequences through averaging.

\subsection{{Computational Complexity of MSV}}

In the Streaming Answers setting, $N$ sequences are decoded in parallel. Assuming delimiters are uniformly distributed throughout a sequence, the total number of delimiters across all sequences—and thus the total number of forward passes through MSV—is $O(NL)$, where $L$ is the average sequence length.
The computational cost of each MSV forward pass is linear in the number of answer tokens it attends to, which is also $O(NL)$. Crucially, the cost is not quadratic because we only need to query the attention mechanism with the tokens in the most recent answer, i.e. we use KV caching (\cref{app:online_inference}). This yields a total MSV computational cost of $O(N^2L^2)$, whereas the decoding of the $N$ sequences itself scales as $O(NL^2)$. Thus, MSV adds an extra factor in $N$, while the dependence on sequence length $L$ matches that of the underlying decoding process.
This is a common issue among methods that incorporate information between sequences, but is usually shown not to be a significant bottleneck for practical values of $N$ \citep{du2023improving,rodionov2025hogwild}.
We provide a simple yet effective solution in \cref{sec:scaling}.

\subsection{{Scaling with MSV M for Large N}}
\label{sec:scaling}

{
To address scalability concerns when $N$ is large (e.g., on the order of hundreds), we employ a simple and effective remedy: use MSV$_M$ with $N$ parallel sequences, where $M$ is kept constant as we scale $N$. This is achieved by partitioning the $N$ sequences into $N/M$ groups, each of size $M$, and processing each group through MSV$_M$ independently. This reduces the computational complexity from $O(N^2L^2)$ to $O(MNL^2)$, or equivalently $O(NL^2)$ when treating $M$ as a constant.
We also note that grouping sequences to reduce computational load is a common strategy in the generative verifier literature, for instance using pairwise knockout rounds \citep{liu2025pairjudge}, or tournaments \citep{toshniwal2025genselect}. We therefore do not claim this to be a novel contribution of our paper.
}

{
We provide the full accuracy-latency tradeoff curves for $N=128$ and $N=256$ in \cref{fig:sa_tradeoff_scaling}, which confirm that MSV$_{64}$ (grouped) consistently dominates the single-sequence baselines across both scales.
More importantly, the plots show that the latency of MSV$_{64}$ scales linearly relative to single-sequence baselines, as we predicted. This confirms that the grouping strategy effectively avoids the potential quadratic blow-up.
}

\input{figure/sa_tradeoff_comp_128/sa_tradeoff_comp_128}

\subsection{{Sensitivity to Delimiter Choice in Streaming Answers}}
\label{app:delimiter_sensitivity}

{
In the Streaming Answers setting, we use delimiter tokens to segment sequences into intermediate answers. While our main experiments use ``Wait'' as the delimiter, we investigate the sensitivity of MSV to this choice by evaluating two alternative delimiters: ``Alternatively'' and ``But'', both of which frequently appear as the first token of a paragraph in mathematical reasoning. On the AIME dataset, ``Wait'', ``Alternatively,'' and ``But'' occur roughly once every 200, 900 and 400 tokens, respectively. We also evaluate a delimiter-free approach that extracts an intermediate answer every 200 tokens, regardless of token content, to confirm that MSV's gains are not contingent on any specific delimiter.
}

\begin{table}[t]
\caption{Area Under the Tradeoff Curve with different delimiters on the AIME dataset with $N=16$ parallel sequences. ``Every 200 tokens'' is a delimiter-free baseline that extracts answers at fixed token intervals.}
\label{tab:delimiter_ablation}
\centering
{
\begin{tabular}{lccc}
\toprule
\textbf{Delimiter} & \textbf{Probe} & \textbf{MSV$_1$} & \textbf{MSV$_{16}$} \\
\midrule
``Wait''          & 1024.2 & 954.8  & \textbf{1078.3} \\
``Alternatively'' & 985.9  & 980.0  & \textbf{1012.4} \\
``But''           & 1039.3 & 1023.6 & \textbf{1070.7} \\
Every 200 tokens  & 910.9  & ---    & \textbf{1013.9} \\
\bottomrule
\end{tabular}
}
\end{table}

{
\cref{tab:delimiter_ablation} reports the area under the tradeoff curve (AUTC) for each delimiter-verifier combination. The results show that while the choice of delimiter does affect the absolute performance of the tradeoff, the relative ordering between verifiers remains consistent across all delimiters: MSV$_{16}$ consistently outperforms Probe across all four settings, including the delimiter-free 200-token interval case, demonstrating that MSV's gains are robust and not contingent on a specific delimiter choice. The delimiter can thus be treated as a standard \emph{hyperparameter}, tunable on a validation set.
}

\subsection{Comparison with GenSelect}
\label{app:genselect}

{
We additionally compare against GenSelect \citep{toshniwal2025genselect}, a generative verifier that prompts an LLM judge to directly select the best solution from a group of $N$ candidates. Since our base model is DeepSeek-R1-Distill-Qwen-1.5B, we use the same model as the summarizer and judge. \cref{tab:genselect} reports best-of-16 accuracy on five datasets, comparing GenSelect against MSV$_{16}$.
}

\begin{table}[t]
\caption{Best-of-16 accuracy comparison between GenSelect and MSV$_{16}$ on five mathematical reasoning benchmarks in the Terminal Answers setting.}
\label{tab:genselect}
\centering
{
\begin{tabular}{lcc}
\toprule
\textbf{Dataset} & \textbf{GenSelect} & \textbf{MSV$_{16}$} \\
\midrule
MATH          & 0.724 & \textbf{0.747} \\
OlympiadBench & 0.500 & \textbf{0.515} \\
AMC12         & 0.455 & \textbf{0.540} \\
AIME          & 0.356 & \textbf{0.462} \\
Omni-MATH     & 0.296 & \textbf{0.320} \\
\bottomrule
\end{tabular}
}
\end{table}

{
MSV$_{16}$ outperforms GenSelect on all five datasets, despite GenSelect employing a full LLM for joint reasoning over candidates. More critically, GenSelect requires approximately 130 seconds on average to judge a group of 16 solutions, compared to 0.006 seconds for MSV$_{16}$ — a roughly 20,000$\times$ difference. This cost makes generative verifiers such as GenSelect fundamentally inapplicable in the \textbf{Streaming Answers} setting, where a judgment must be issued at every intermediate answer across the decoding process.
Note that the 0.006 seconds for MSV$_{16}$ is for Terminal Answers settings, hence why it differs from the streaming answers latency reported in \cref{tab:streaming_verifier_latency} which is measured across many intermediate answers.
}

\subsection{{Training Set Size}}
\label{app:training_set_size}

Our training set consists of 224 problems sampled from the DeepMath-103K training set. While some readers may be concerned about how small this is, we sample 64 model responses per problem, yielding $224 \times 64 = 14{,}336$ sequences in total.
To further validate the choice of our training set size, we conducted preliminary experiments with Probe in the {Terminal Answers} setting, varying the number of training problems. \cref{tab:training_size} shows that performance plateaus beyond 224 problems, with doubling to 448 problems yielding no improvement. This saturation suggests that our verifiers efficiently learn the verification task with relatively few problems, which represents a practical advantage: MSV can be trained quickly without requiring extensive problem collections or prolonged training procedures.
This is also consistent with \citet{zhang2025reasoning} who report that a training set of 1,000 sequences was enough for Probe to generalize.

\begin{table}[t]
\caption{Brier score of Probe on the AIME dataset as a function of training set size}
\label{tab:training_size}
\centering
{
\begin{tabular}{lc}
\toprule
\textbf{\# of problems} & \textbf{Brier score} \\
\midrule
56  & 0.0833 \\
112 & 0.0821 \\
224 & 0.0809 \\
448 & 0.0812 \\
\bottomrule
\end{tabular}
}
\end{table}

\subsection{{MSV with Mean Across Answer Tokens}}
\label{sec:mean_tokens}

{
In our main experiments, MSV uses only transformer block output at the last token position of each answer in order to obtain the logit. This design choice was inherited from prior work on probes \citep{zhang2025reasoning}, which operate exclusively on the final hidden state. We hypothesized that the attention layer would be sufficiently expressive to aggregate relevant information from all answer tokens.
}

{
However, an alternative design is to use the mean of hidden states across all answer tokens instead of only the last token. To evaluate this variant, we compare MSV$_N$ against MSV$_N$ (Mean) in the Terminal Answers setting on the AIME dataset.
}

\begin{table}[t]
\caption{Brier score comparison between MSV$_N$ using the last token versus the mean of answer tokens on the AIME dataset in the Terminal Answers setting.}
\label{tab:mean_tokens}
\centering
{
\begin{tabular}{lcccc}
\toprule
\textbf{Method} & \textbf{$N=1$} & \textbf{$N=4$} & \textbf{$N=16$} & \textbf{$N=64$} \\
\midrule
MSV$_N$         & 0.0819 & 0.0687 & 0.0572 & 0.0387 \\
MSV$_N$ (Mean)  & 0.0756 & 0.0641 & 0.0544 & 0.0382 \\
\bottomrule
\end{tabular}
}
\end{table}

{
\cref{tab:mean_tokens} shows that the mean variant provides a modest but consistent improvement in Brier score across all values of $N$. However, this improvement comes at a computational cost: the mean variant requires querying the attention mechanism with all answer tokens rather than just the last token, resulting in slightly higher computational overhead. Given this tradeoff and the already strong performance of the last-token variant, we use the last-token approach in our main experiments for computational efficiency.
}

%% file: table/main_ablation.tex
\begin{table}[t]
\caption{Ablation of MSV architectural choices in the Streaming Answers setting. We report calibration metrics on the AIME dataset.}
\label{tab:main_ablation}
\centering
\begin{tabular}{lccc}
\toprule
\textbf{Method} & \textbf{AUROC $\uparrow$} & \textbf{Brier $\downarrow$} & \textbf{NLL $\downarrow$} \\
\midrule
MSV$_1$ & 0.906\PM{0.008} & 0.064\PM{0.003} & 0.241\PM{0.011} \\
MSV$_{16}$ & \textbf{0.953}\PM{0.007} & \textbf{0.038}\PM{0.003} & \textbf{0.150}\PM{0.011} \\
Ablation & 0.936\PM{0.0016} & 0.047\PM{0.009} & 0.190\PM{0.0333} \\
\bottomrule
\end{tabular}
\end{table}



%% file: figure/sa_tradeoff_comp_128/sa_tradeoff_comp_128.tex
\begin{figure*}[t]
    \centering
    \begin{minipage}[t]{0.49\linewidth}
        \centering
        \includegraphics[width=\linewidth]{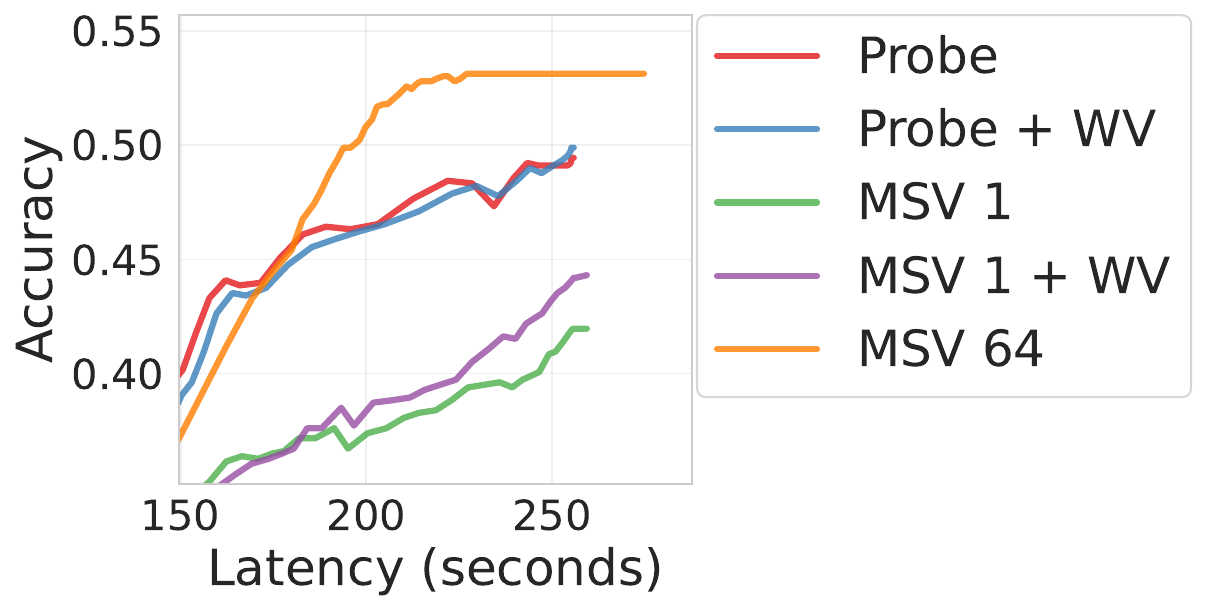}
        \subcaption{$N=128$}
        \label{fig:sa_tradeoff_comp_128}
    \end{minipage}
    \hfill
    \begin{minipage}[t]{0.49\linewidth}
        \centering
        \includegraphics[width=\linewidth]{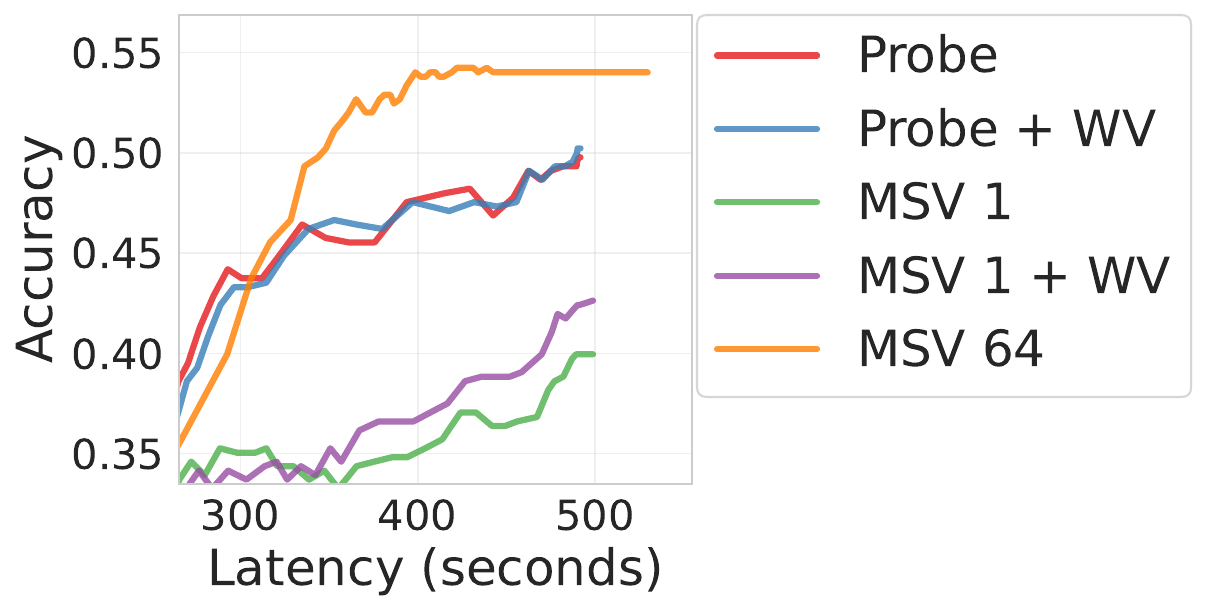}
        \subcaption{$N=256$}
        \label{fig:sa_tradeoff_comp_256}
    \end{minipage}
    \caption{Accuracy-latency tradeoff curves in \textbf{Streaming Answers} on AIME. We use the grouping strategy with MSV$_{64}$ for large $N$.}
    \label{fig:sa_tradeoff_scaling}
\end{figure*}

%% file: appendix/method_details.tex
\section{Details on Method}

\subsection{Multi-Sequence Verifier Pseudocode}
\label{app:msv_algorithm}

\input{algorithm/msv}

In \cref{alg:msv}, we present a pseudocode of MSV's forward pass. Note that \cref{alg:msv} presents the \emph{offline} version of the forward pass, which recomputes $Z^{(t)}$ from all accumulated answer tokens at each readout time $t$. In practice, we use a more efficient \emph{online} version during inference, described next.

\subsection{Online Inference with KV Caching}
\label{app:online_inference}

\cref{alg:msv} processes all answers accumulated up to time $t$ in a single forward pass, which would require recomputing the full attention outputs from scratch each time a new answer arrives. In practice, we exploit the causal structure of the attention masks in the \textbf{Streaming Answers} setting to avoid redundant computation via key-value (KV) caching, analogously to how autoregressive language model inference reuses cached states from prior tokens.

Because the causal constraint in the \textbf{Streaming Answers} setting means that an answer $a_k^{(n)}$ may only attend to earlier answers $a_{k'}^{(m)}$ with $\tau_{k'}^{(m)} + L_{k'}^{(m)} \leq \tau_k^{(n)} + L_k^{(n)}$, the keys and values corresponding to all previously generated answers remain unchanged when a new answer arrives. Their attention outputs therefore do not need to be recomputed.

\paragraph{Online procedure.}
When a new intermediate answer $a_k^{(n)}$ (with $L_k^{(n)}$ tokens) is generated at time $\tau_k^{(n)}$, the online algorithm proceeds as follows:
\begin{enumerate}[nosep]
  \item \textbf{Query only the new tokens.} Compute query vectors for the $L_k^{(n)}$ new tokens only, while reading key and value vectors for \emph{all} tokens (new and cached) to form the attention output for the new tokens. KV vectors for the new tokens are appended to the cache.
  \item \textbf{Extract the last-token output.} Since only the last token's representation $z_{k,L_k^{(n)}}^{(n)}$ is needed for prediction (the attention layer has already aggregated within-answer information), we pass only this representation to the MLP and prediction head.
  \item \textbf{Update $\gamma_k^{(n)}$ and predict.} Compute $\gamma_k^{(n)}$ using the causal definition (\cref{sec:method}), augment the last-token representation, and output the logit $\tilde{y}_k^{(n)}$.
\end{enumerate}

\paragraph{Complexity.}
Each per-answer forward pass in the online algorithm attends $L_k^{(n)}$ query tokens against a key-value cache of size $T_{\mathrm{acc}}$ (the total number of answer tokens accumulated so far), giving an attention cost of $O(L_k^{(n)} \cdot T_{\mathrm{acc}})$ per new answer, compared to $O(T_{\mathrm{acc}}^2)$ for a full recomputation. The MLP and feature augmentation steps operate on a single token and are $O(d)$.

\paragraph{Equivalence to the offline algorithm.}
The online and offline algorithms produce numerically identical outputs. This follows from the causal masking: the attention output for any token $u$ depends only on tokens $v$ with $\tau(v) \leq \tau(u)$, so processing tokens in arrival order with a growing KV cache yields the same result as processing the full sequence at once. We have verified this numerically in our implementation. The latency measurements reported in \cref{app:latency_analysis} use the online algorithm; the accuracy experiments use the offline algorithm (which is simpler to implement for batch evaluation over pre-collected sequences).

%% file: algorithm/msv.tex
\makeatletter
\def\ALG@special@indent{%
  \ifdim\ALG@thistlm=0pt\relax
    \hskip-\leftmargin
  \else
    \hskip\ALG@thistlm
  \fi
}
\newcommand{\TrainableParameters}[1]{%
  \item[]\noindent\ALG@special@indent
  \textbf{Trainable parameters}: #1%
}
\makeatother

\begin{algorithm}[t]
\caption{{Multi-Sequence Verifier (MSV) Forward Pass}}
\label{alg:msv}
{
\begin{algorithmic}[1]
\Require $N$ sequences with hidden states $\{h^{(n)}_{k,i}\}$, readout time $t$
\Ensure Correctness predictions $\{\tilde{y}^{(n)}_k\}$ for all answers up to time $t$
\TrainableParameters{Sequence embeddings $\{\mathsf{e}^{(n)}\}$, QKV projections $(W_Q, W_K, W_V)$, mask weights $\{\mathbf{w}_h\}$, MLP \& LN, feature MLP, prediction head $(\mathbf{w}, b)$}
\State $U^{(t)} \gets \text{concat}\left(\left\{\left\{\{h^{(n)}_{k,i} + \mathsf{e}^{(n)}\}_{i=1}^{L^{(n)}_k}\right\}_{\tau^{(n)}_k + L_k\nth \leq t}\right\}_{n=1}^N\right)$
\State Compute query, key, value projections: $(Q^{(t)}_h, K^{(t)}_h, V^{(t)}_h)_{h=1}^H$
\For{each head $h=1,\ldots,H$ and mask $j=1,\ldots,J$}
  \State $A^{(t)}_{h,j} \gets \text{softmax}\left(\frac{Q^{(t)}_h(K^{(t)}_h)^\top}{\sqrt{d}} + \log M_j\right)V^{(t)}_h$
\EndFor
\State $\tilde{U}^{(t)} \gets \sum_{h=1}^H \sum_{j=1}^J \alpha_{h,j} A^{(t)}_{h,j}$ \quad where $(\alpha_{h,1},\ldots,\alpha_{h,J}) = \text{softmax}(\mathbf{w}_h)$
\State $Z^{(t)} \gets (U^{(t)} + \tilde{U}^{(t)}) + \text{MLP}(\text{LN}(U^{(t)} + \tilde{U}^{(t)}))$
\For{each answer $a^{(n)}_k$ generated by time $t$}
  \If{Terminal Answers}
    \State $S^{(n)}_k \gets \{1, \dots, N\}$
    \For{$m\!\in\!S^{(n)}_k$}
      \State $j_m \gets 1$
    \EndFor
  \ElsIf{Streaming Answers}
    \State $S^{(n)}_k \gets \{m : \tau^{(m)}_1 + L^{(m)}_1 \leq \tau^{(n)}_k + L^{(n)}_k\}$
    \For{$m\!\in\!S^{(n)}_k$}
      \State $j_m \gets \max\{j : \tau^{(m)}_j + L^{(m)}_j \leq \tau^{(n)}_k + L^{(n)}_k\}$
    \EndFor
  \EndIf
  \State $\gamma^{(n)}_k \gets \dfrac{1}{|S^{(n)}_k|}\displaystyle\sum_{m \in S^{(n)}_k} \mathbf{1}[a^{(m)}_{j_m} \sim a^{(n)}_k]$
  \State $\bar{z}^{(n)}_k \gets z^{(n)}_{k,L^{(n)}_k} + \text{MLP}(\gamma^{(n)}_k)$
\EndFor
\If{Terminal Answers}
  \State $\tilde{y}^{(n)}_1 \gets \sigma\left(\frac{1}{|\mathcal{C}(n)|}\sum_{m \in \mathcal{C}(n)}(\mathbf{w}^\top \bar{z}^{(m)}_1 + b)\right)$ for all $n$
\ElsIf{Streaming Answers}
  \State $\tilde{y}^{(n)}_k \gets \sigma(\mathbf{w}^\top \bar{z}^{(n)}_k + b)$ for all $(n,k)$
\EndIf
\State \Return $\{\tilde{y}^{(n)}_k\}$
\end{algorithmic}
}
\end{algorithm}

%% file: appendix/full_report.tex
\section{Full Report of Main Experiments}
\label{app:full_report}

This section presents the full results on our main experiments, including all benchmarks, evaluation metrics, and baselines including the training-free methods. All tables and figures are provided here in complete form for reference.

\subsection{Full report on the \textbf{Terminal Answers} setting.}
\label{app:full_report_ta}

\cref{tab:ta_auroc_brier} and \cref{tab:ta_ece_nll} report all the standard calibration metrics--AUROC, BS, ECE, and NLL--in the \textbf{Terminal Answers} setting, including training-free baselines (Token Probs, Self-consistency). We find that MSV$_N$ generally improves over the baselines with increasing $N$, following the trend reported in \cref{subsec:terminal}.
Similarly, \cref{tab:acc_bon_only_reordered_terminal} reports the full best-of-N accuracy across all values of $N$. We find that the best single-sequence baselines (Probe+WV$_4$) perform slightly better than MSV$_4$ at $N=4$, and that MSV$_{N}$ clearly and consistently improves over all baselines at $N=16$ and $N=64$. This is also consistent with our reports in \cref{subsec:terminal}. Finally, we show all the calibration metrics of best-of-N confidence, namely ECE and BS, in \cref{tab:calib_bon_only_reordered_terminal}. In terms of best-of-N calibration, MSV$_N$ outperforms the best single-sequence baselines across all values of $N$, implying much more reliable confidence levels.

\input{table/ta_standard}

\input{table/ta_bon}

\clearpage

\subsection{Full report on the \textbf{Streaming Answers} setting.}
\label{app:full_report_sa}

Next, we analyze the full results for the \textbf{Streaming Answers} setting. \cref{tab:sa_standard_auroc_brier} and \cref{tab:sa_standard_ece_nll} show the standard calibration metrics.
Consistent with the findings in \cref{sub:streaming_answer}, the results demonstrate that MSV$_{N}$ generally and significantly outperforms single-sequence baselines across most metrics and datasets, with performance scaling with $N$. Minor exceptions exist: for instance, MSV$_{64}$ slightly deteriorates compared to MSV$_{16}$ on some metrics (e.g., OlympiadBench AUROC, AMC12 Brier).
Token Probs performs poorly across all datasets, so we omit them from all parallel early stopping experiments.

\input{table/sa_standard}

\begin{figure*}[t]
    \centering
    \includegraphics[width=0.98\linewidth]{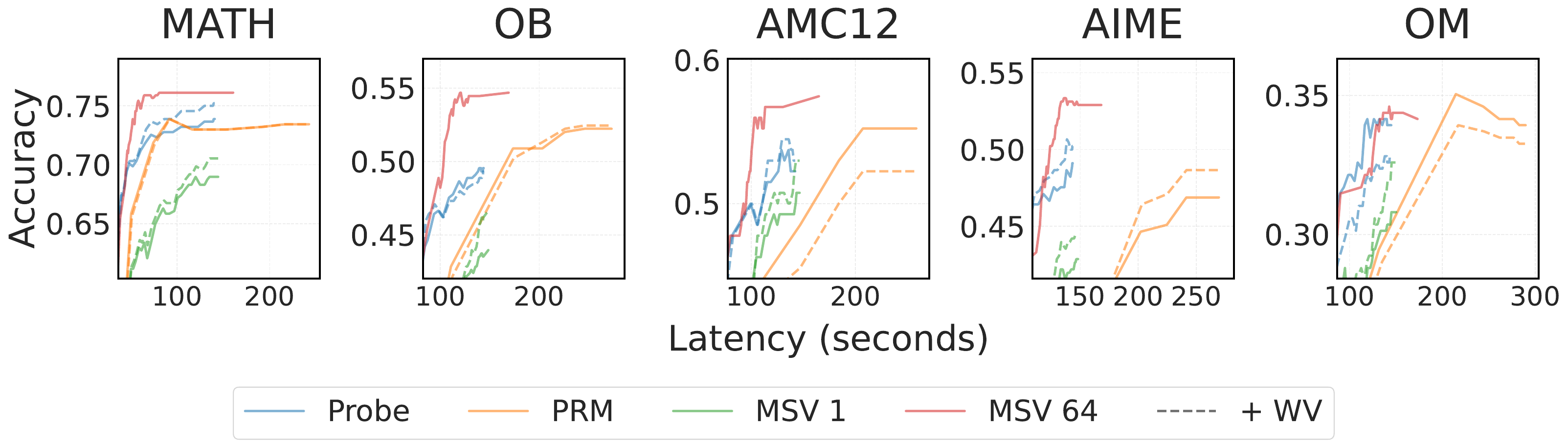}
    \caption{Accuracy-latency tradeoff curves in \textbf{Streaming Answers}, with $N=64$. Curves that lie higher are superior in that they require lower latency to achieve the same accuracy. Accuracy-token tradeoff curves are in \cref{fig:sa_tradeoff_token_position}.}
    \label{fig:sa_tradeoff_comp}
\end{figure*}

\begin{figure*}[t]
    \centering
    \begin{subfigure}{\linewidth}
        \includegraphics[width=0.98\linewidth]{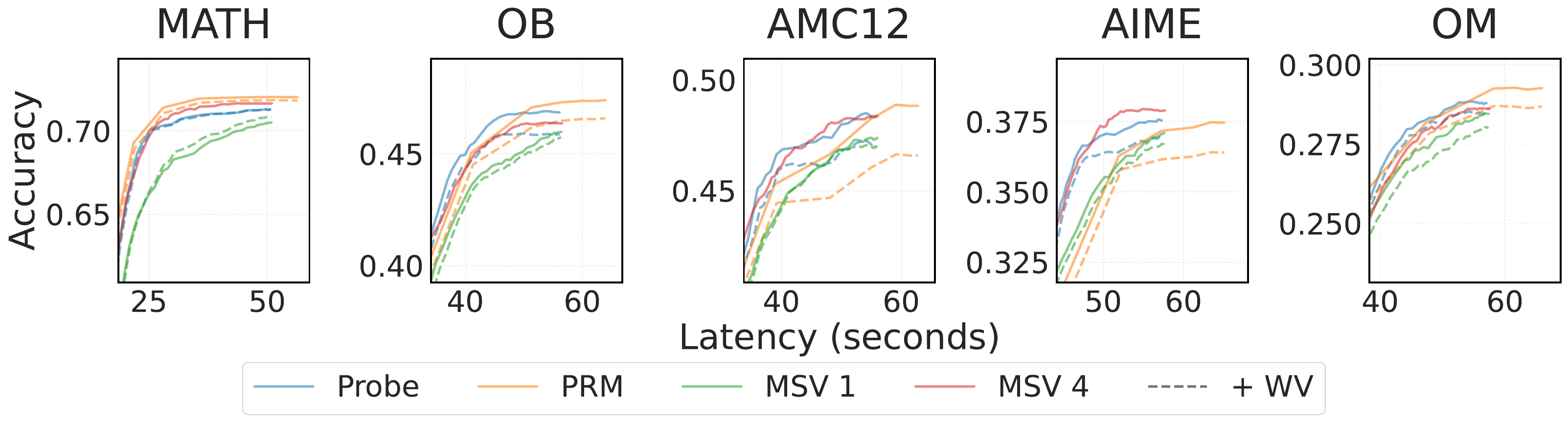}
        \caption{$N=4$}
    \end{subfigure}
    \begin{subfigure}{\linewidth}
        \includegraphics[width=0.98\linewidth]{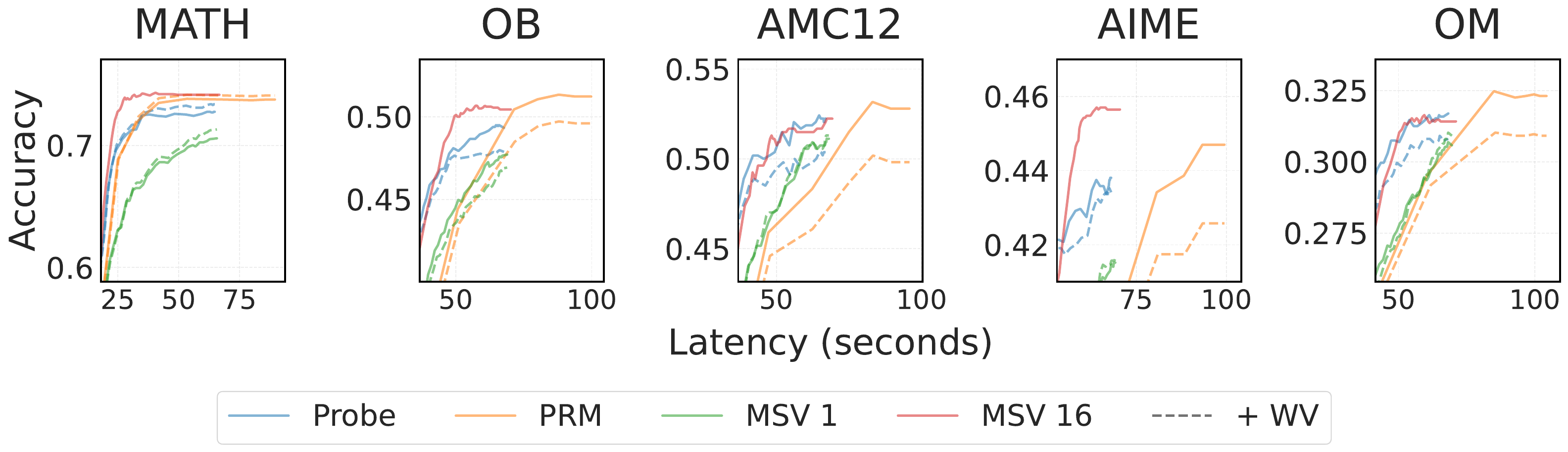}
        \caption{$N=16$}
    \end{subfigure}
    \caption{Accuracy-latency tradeoff curves in \textbf{Streaming Answers}, for $N\in\{4,16\}$.}
    \label{fig:sa_tradeoff_comp_small_n}
\end{figure*}

We provide the full accuracy-latency tradeoff curves for $N=64$ in \cref{fig:sa_tradeoff_comp}, and for $N=4$ and $N=16$ in \cref{fig:sa_tradeoff_comp_small_n}.
Consistent with the \textbf{Terminal Answers} setting, the advantage of MSV$_N$ over single-sequence baselines becomes more pronounced starting at $N=16$. To be concrete, MSV$_{16}$ improves the tradeoff on MATH, OB, and AIME, while it is on par with the best baseline Probe on AMC12 and OM.

\clearpage

\subsection{Full Report of Robustness to Equivalence Checker Quality}
\label{sec:full_report_checker_robustness}

\paragraph{Experimental details.}
We experiment with five alternative checkers of decreasing quality---exact string match, Jaccard similarity ($\geq 0.8$), embedding cosine similarity at two thresholds ($\geq 0.85$ and $\geq 0.6$), and random coin flip---while always using SymPy as the oracle for measuring correctness.
We use the \texttt{bge-small-en-v1.5} model \citep{xiao2024c} for embedding cosine similarity.
For MSV, we remove logit averaging in \cref{eq:logit_avg}, so the weak equivalence checkers are used only within Multi-Mask Transformer Block and feature augmentation, both of which are trained components.
We also experiment with MSV$_{64}$ that's been trained on SymPy and with logit averaging, but run using the weak checkers.

\paragraph{Results.}
We report in \cref{tab:full_checker_robustness} the full results. ``Checker accuracy'' is the accuracy of the weak checker in classifying whether two independently sampled answers on a problem are equivalent according to SymPy.
First, we find that removing logit averaging drops the best-of-64 accuracy of MSV$_{64}$ from 0.356 to 0.348, when evaluated using the ground truth checker SymPy.
However, MSV$_{64}$ trained on SymPy with logit averaging quickly degrades in accuracy when run with checkers of decreasing quality.
This is expected due to the discrepancy between the checker used during training and the checker used during evaluation.
Self-consistency and weighted voting also quickly degrade to the average accuracy of the base language model.
On the other hand, MSV$_{64}$ degrades gracefully with checkers of decreasing quality, because it has been trained to use the weak checkers.
This shows the importance of end-to-end training in MSV$_{64}$, and highlights its robustness.

\begin{table}[t]
\caption{Best-of-64 accuracy on Omni-MATH under different equivalence checkers. Checker accuracy is the pairwise agreement with SymPy. Answer correctness is always evaluated with SymPy.}
\label{tab:full_checker_robustness}
\centering
\setlength{\tabcolsep}{2pt}
\renewcommand{\arraystretch}{0.95}
\small
\begin{adjustbox}{max width=\textwidth}
\begin{tabular}{lccccccc}
\toprule
\textbf{Checker} & \makecell{\textbf{Checker}\\\textbf{accuracy}} & \makecell{\textbf{Self-}\\\textbf{consistency}} & \textbf{Probe+WV} & \textbf{PRM+WV} & \textbf{MSV$_1$+WV} & \makecell{\textbf{MSV$_{64}$}\\\textbf{(SymPy)}} & \makecell{\textbf{MSV$_{64}$}\\\textbf{(w/o logit avg.)}} \\
\midrule
SymPy (reference)              & 100.0\%              & 0.279\PM{0.000} & 0.321\PM{0.002} & 0.297\PM{0.000} & 0.318\PM{0.002} & \textbf{0.356}\PM{0.002} & 0.348\PM{0.007} \\
Exact string match             & \phantom{0}99.0\%    & 0.277\PM{0.000} & 0.321\PM{0.002} & 0.295\PM{0.000} & 0.318\PM{0.002} & \textbf{0.355}\PM{0.003} & 0.345\PM{0.005} \\
Jaccard ($\geq 0.8$)           & \phantom{0}97.0\%    & 0.277\PM{0.000} & 0.316\PM{0.002} & 0.292\PM{0.000} & 0.315\PM{0.003} & \textbf{0.352}\PM{0.002} & 0.345\PM{0.008} \\
Embedding ($\geq 0.85$)        & \phantom{0}89.4\%    & 0.272\PM{0.000} & 0.301\PM{0.002} & 0.286\PM{0.000} & 0.298\PM{0.003} & 0.328\PM{0.002} & \textbf{0.344}\PM{0.008} \\
Embedding ($\geq 0.6$)         & \phantom{0}61.2\%    & 0.257\PM{0.000} & 0.264\PM{0.003} & 0.257\PM{0.000} & 0.263\PM{0.002} & 0.256\PM{0.001} & \textbf{0.338}\PM{0.009} \\
Random                         & \phantom{0}46.7\%    & 0.246\PM{0.000} & 0.246\PM{0.001} & 0.243\PM{0.000} & 0.246\PM{0.001} & 0.251\PM{0.003} & \textbf{0.336}\PM{0.015} \\
\bottomrule
\end{tabular}
\end{adjustbox}
\end{table}

\subsection{{Full Report of Experiment with Different Base Language Models}}
\label{app:base_lm_generalization}

{
We evaluate the robustness of \gls{msv} across different base language models. Concretely, we consider three additional base LMs: DeepSeek-R1-Distill-Llama-8B, Qwen3-1.7B in thinking mode, and Llama-3.2-1B-Instruct.
\cref{tab:base_lm_calib} reports calibration metrics for different verifiers, and \cref{tab:base_lm_bo64} reports best-of-64 accuracy.
Unless otherwise stated, we reuse the same training setup and hyperparameters as in the main experiments.
}

\begin{table}[t]
\caption{Brier score, AUROC, ECE, and NLL of verifiers across different base language models. R1-8B is abbreviation for DeepSeek-R1-Distill-Llama-8B.}
\label{tab:base_lm_calib}
\centering
\setlength{\tabcolsep}{3pt}
\renewcommand{\arraystretch}{0.95}
{
\begin{tabular}{lllccc}
\toprule
\textbf{Base LM} & \textbf{Dataset} & \textbf{Metric} & \textbf{Probe} & \textbf{MSV$_1$} & \textbf{MSV$_{64}$} \\
\midrule
\multirow{4}{*}{R1-8B} & \multirow{4}{*}{AIME}
  & Brier $\downarrow$ & 0.0673\PM{0.0104} & 0.0738\PM{0.0151} & \textbf{0.0363}\PM{0.0090} \\
  & & AUROC $\uparrow$   & 0.9672\PM{0.0017} & 0.9750\PM{0.0043} & \textbf{0.9901}\PM{0.0012} \\
  & & ECE $\downarrow$   & 0.0479\PM{0.0272} & 0.0694\PM{0.0201} & \textbf{0.0367}\PM{0.0148} \\
  & & NLL $\downarrow$   & 0.2555\PM{0.0407} & 0.3240\PM{0.0708} & \textbf{0.1868}\PM{0.0352} \\
\midrule
\multirow{4}{*}{Qwen3-1.7B} & \multirow{4}{*}{AIME}
  & Brier $\downarrow$ & 0.0848\PM{0.0333} & 0.0770\PM{0.0059} & \textbf{0.0509}\PM{0.0246} \\
  & & AUROC $\uparrow$   & 0.9530\PM{0.0174} & 0.9557\PM{0.0053} & \textbf{0.9863}\PM{0.0011} \\
  & & ECE $\downarrow$   & 0.0627\PM{0.0656} & \textbf{0.0438}\PM{0.0169} & 0.0639\PM{0.0448} \\
  & & NLL $\downarrow$   & 0.2884\PM{0.1024} & 0.2705\PM{0.0149} & \textbf{0.1945}\PM{0.0722} \\
\midrule
\multirow{4}{*}{Llama-3.2-1B} & \multirow{4}{*}{MATH}
  & Brier $\downarrow$ & 0.1368\PM{0.0363} & 0.1324\PM{0.0050} & \textbf{0.0565}\PM{0.0056} \\
  & & AUROC $\uparrow$   & 0.7856\PM{0.0137} & 0.7294\PM{0.0304} & \textbf{0.9513}\PM{0.0010} \\
  & & ECE $\downarrow$   & 0.0982\PM{0.1043} & 0.0936\PM{0.0157} & \textbf{0.0430}\PM{0.0159} \\
  & & NLL $\downarrow$   & 0.4264\PM{0.0904} & 0.4931\PM{0.0732} & \textbf{0.2043}\PM{0.0150} \\
\bottomrule
\end{tabular}
}
\end{table}

\begin{table}[t]
\caption{Best-of-64 accuracy across different base language models. R1-8B is abbreviation for DeepSeek-R1-Distill-Llama-8B.}
\label{tab:base_lm_bo64}
\centering
\setlength{\tabcolsep}{4pt}
\renewcommand{\arraystretch}{1.08}
{
\begin{tabular}{llccccc}
\toprule
\textbf{Base LM} & \textbf{Dataset} & \textbf{Probe} & \textbf{Probe+WV} & \textbf{MSV$_1$} & \textbf{MSV$_1$+WV} & \textbf{MSV$_{64}$} \\
\midrule
R1-8B & AIME & 0.6116\PM{0.0098} & 0.6357\PM{0.0131} & 0.6047\PM{0.0088} & 0.6492\PM{0.0091} & \textbf{0.6625}\PM{0.0054} \\
Qwen3-1.7B & AIME & 0.5223\PM{0.0130} & 0.5089\PM{0.0179} & 0.4545\PM{0.0097} & 0.5288\PM{0.0089} & \textbf{0.5357}\PM{0.0109} \\
Llama-3.2-1B & MATH & 0.2393\PM{0.0143} & 0.3375\PM{0.0051} & 0.1670\PM{0.0201} & 0.2946\PM{0.0192} & \textbf{0.3777}\PM{0.0051} \\
\bottomrule
\end{tabular}
}
\end{table}

{
\paragraph{DeepSeek-R1-Distill-Llama-8B on AIME.}
We first consider a larger base model, DeepSeek-R1-Distill-Llama-8B, in the AIME Terminal Answers setup.
Increasing $N$ consistently improves the calibration of MSV, and MSV$_{64}$ yields the highest AUROC and lowest Brier score.
When used for best-of-64 selection, MSV$_{64}$ also outperforms all baselines, including Probe with and without weighted voting.
}

{
\paragraph{Qwen3-1.7B (thinking mode) on AIME.}
We next evaluate Qwen3-1.7B in thinking mode on AIME.
MSV achieves progressively better calibration as $N$ increases, with substantial gains in both Brier score and AUROC compared to Probe.
MSV$_{64}$ also improves best-of-64 accuracy over all baselines.
}

{
\paragraph{Llama-3.2-1B-Instruct on MATH.}
Finally, we consider Llama-3.2-1B-Instruct, a non-reasoning model whose accuracy on AIME is close to zero.
For this model we therefore train and evaluate on the MATH train/test splits.
Even in this lower-accuracy regime, MSV$_{16}$ and MSV$_{64}$ significantly improve calibration over Probe, and MSV$_{64}$ achieves the best best-of-64 accuracy.
}

{
Overall, these results indicate that the benefits of \gls{msv} are not restricted to a particular base model: across all three additional base LMs, MSV systematically improves both calibration and best-of-N accuracy.
}

%% file: table/ta_standard.tex
\begin{table}[t]
\caption{AUROC ($\uparrow$) and Brier score ($\downarrow$) of verifiers on terminal answers.}
\label{tab:ta_auroc_brier}
\setlength{\tabcolsep}{3pt}
\renewcommand{\arraystretch}{0.9}
\centering
\resizebox{\textwidth}{!}{
\begin{tabular}{lrrrrrrrrrr}
\toprule
& \multicolumn{2}{c}{\textbf{MATH}}
& \multicolumn{2}{c}{\textbf{OlympiadBench}}
& \multicolumn{2}{c}{\textbf{AMC12}}
& \multicolumn{2}{c}{\textbf{AIME}}
& \multicolumn{2}{c}{\textbf{Omni\text{-}MATH}} \\
\cmidrule(lr){2-3}\cmidrule(lr){4-5}\cmidrule(lr){6-7}\cmidrule(lr){8-9}\cmidrule(lr){10-11}
\multicolumn{1}{c}{\textbf{Method}}
& \multicolumn{1}{c}{\textbf{AUROC $\uparrow$}} & \multicolumn{1}{c}{\textbf{Brier $\downarrow$}}
& \multicolumn{1}{c}{\textbf{AUROC $\uparrow$}} & \multicolumn{1}{c}{\textbf{Brier $\downarrow$}}
& \multicolumn{1}{c}{\textbf{AUROC $\uparrow$}} & \multicolumn{1}{c}{\textbf{Brier $\downarrow$}}
& \multicolumn{1}{c}{\textbf{AUROC $\uparrow$}} & \multicolumn{1}{c}{\textbf{Brier $\downarrow$}}
& \multicolumn{1}{c}{\textbf{AUROC $\uparrow$}} & \multicolumn{1}{c}{\textbf{Brier $\downarrow$}} \\
\midrule
Token Probs
& 0.5320\PM{0.0000} & 0.2504\PM{0.0000}
& 0.7704\PM{0.0000} & 0.3696\PM{0.0000}
& 0.7350\PM{0.0000} & 0.3540\PM{0.0000}
& 0.8999\PM{0.0000} & 0.3073\PM{0.0000}
& 0.7814\PM{0.0000} & 0.4467\PM{0.0000} \\
Probe
& 0.8790\PM{0.0052} & 0.1346\PM{0.0107}
& 0.9022\PM{0.0038} & 0.1271\PM{0.0144}
& 0.8994\PM{0.0062} & 0.1438\PM{0.0214}
& 0.9461\PM{0.0034} & 0.0859\PM{0.0147}
& 0.8836\PM{0.0008} & 0.1340\PM{0.0252} \\
\gls{msv} 1
& 0.8566\PM{0.0166} & 0.1460\PM{0.0063}
& 0.8926\PM{0.0048} & 0.1310\PM{0.0061}
& 0.8890\PM{0.0080} & 0.1498\PM{0.0099}
& 0.9469\PM{0.0025} & 0.0819\PM{0.0066}
& 0.8711\PM{0.0062} & 0.1460\PM{0.0103} \\
PRM
& 0.8156\PM{0.0000} & 0.1747\PM{0.0000}
& 0.8958\PM{0.0000} & 0.1689\PM{0.0000}
& 0.8521\PM{0.0000} & 0.1807\PM{0.0000}
& 0.9374\PM{0.0000} & 0.1472\PM{0.0000}
& 0.8826\PM{0.0000} & 0.1584\PM{0.0000} \\
\midrule
Token Probs + WV$_4$
& 0.7922\PM{0.0000} & 0.1678\PM{0.0000}
& 0.8392\PM{0.0000} & 0.2166\PM{0.0000}
& 0.7978\PM{0.0000} & 0.2448\PM{0.0000}
& 0.8223\PM{0.0000} & 0.2239\PM{0.0000}
& 0.8085\PM{0.0000} & 0.2620\PM{0.0000} \\
Self-consistency$_4$
& 0.7876\PM{0.0000} & 0.1686\PM{0.0000}
& 0.8290\PM{0.0000} & 0.2180\PM{0.0000}
& 0.7893\PM{0.0000} & 0.2477\PM{0.0000}
& 0.8001\PM{0.0000} & 0.2274\PM{0.0000}
& 0.7986\PM{0.0000} & 0.2627\PM{0.0000} \\
Probe + WV$_4$
& 0.7930\PM{0.0033} & 0.1649\PM{0.0008}
& 0.8494\PM{0.0018} & 0.2164\PM{0.0017}
& 0.8222\PM{0.0040} & 0.2367\PM{0.0010}
& 0.8624\PM{0.0037} & 0.2100\PM{0.0017}
& 0.8251\PM{0.0025} & 0.2704\PM{0.0028} \\
\gls{msv} 1 + WV$_4$
& 0.7920\PM{0.0023} & 0.1656\PM{0.0005}
& 0.8479\PM{0.0009} & 0.2168\PM{0.0013}
& 0.8204\PM{0.0026} & 0.2371\PM{0.0014}
& 0.8636\PM{0.0012} & 0.2096\PM{0.0008}
& 0.8219\PM{0.0012} & 0.2711\PM{0.0021} \\
PRM + WV$_4$
& 0.7942\PM{0.0000} & 0.1653\PM{0.0000}
& 0.8491\PM{0.0000} & 0.2126\PM{0.0000}
& 0.8136\PM{0.0000} & 0.2403\PM{0.0000}
& 0.8444\PM{0.0000} & 0.2171\PM{0.0000}
& 0.8200\PM{0.0000} & 0.2602\PM{0.0000} \\
\gls{msv} 4
& \textbf{0.8945}\PM{0.0135} & \textbf{0.1348}\PM{0.0055}
& \textbf{0.9298}\PM{0.0036} & \textbf{0.1108}\PM{0.0064}
& \textbf{0.9125}\PM{0.0127} & \textbf{0.1406}\PM{0.0123}
& \textbf{0.9628}\PM{0.0040} & \textbf{0.0687}\PM{0.0132}
& \textbf{0.8986}\PM{0.0025} & \textbf{0.1296}\PM{0.0128} \\
\midrule
Token Probs + WV$_{16}$
& 0.8155\PM{0.0000} & 0.1474\PM{0.0000}
& 0.8801\PM{0.0000} & 0.1690\PM{0.0000}
& 0.8364\PM{0.0000} & 0.1961\PM{0.0000}
& 0.8531\PM{0.0000} & 0.1662\PM{0.0000}
& 0.8457\PM{0.0000} & 0.1968\PM{0.0000} \\
Self-consistency$_{16}$
& 0.8149\PM{0.0000} & 0.1482\PM{0.0000}
& 0.8751\PM{0.0000} & 0.1704\PM{0.0000}
& 0.8299\PM{0.0000} & 0.1991\PM{0.0000}
& 0.8422\PM{0.0000} & 0.1696\PM{0.0000}
& 0.8398\PM{0.0000} & 0.1976\PM{0.0000} \\
Probe + WV$_{16}$
& 0.8228\PM{0.0018} & 0.1416\PM{0.0007}
& 0.9028\PM{0.0022} & 0.1588\PM{0.0010}
& 0.8728\PM{0.0030} & 0.1759\PM{0.0025}
& 0.9092\PM{0.0055} & 0.1368\PM{0.0039}
& 0.8714\PM{0.0026} & 0.1921\PM{0.0006} \\
\gls{msv} 1 + WV$_{16}$
& 0.8228\PM{0.0033} & 0.1425\PM{0.0003}
& 0.9008\PM{0.0010} & 0.1612\PM{0.0008}
& 0.8691\PM{0.0029} & 0.1786\PM{0.0018}
& 0.9057\PM{0.0024} & 0.1404\PM{0.0015}
& 0.8672\PM{0.0010} & 0.1946\PM{0.0008} \\
PRM + WV$_{16}$
& 0.8149\PM{0.0000} & 0.1457\PM{0.0000}
& 0.8888\PM{0.0000} & 0.1647\PM{0.0000}
& 0.8466\PM{0.0000} & 0.1922\PM{0.0000}
& 0.8693\PM{0.0000} & 0.1588\PM{0.0000}
& 0.8524\PM{0.0000} & 0.1953\PM{0.0000} \\
\gls{msv} 16
& \textbf{0.9062}\PM{0.0182} & \textbf{0.1255}\PM{0.0038}
& \textbf{0.9469}\PM{0.0021} & \textbf{0.1005}\PM{0.0064}
& \textbf{0.9231}\PM{0.0144} & \textbf{0.1315}\PM{0.0095}
& \textbf{0.9732}\PM{0.0015} & \textbf{0.0572}\PM{0.0114}
& \textbf{0.9109}\PM{0.0034} & \textbf{0.1258}\PM{0.0113} \\
\midrule
Token Probs + WV$_{64}$
& 0.8304\PM{0.0000} & 0.1432\PM{0.0000}
& 0.8901\PM{0.0000} & 0.1591\PM{0.0000}
& 0.8454\PM{0.0000} & 0.1871\PM{0.0000}
& 0.8608\PM{0.0000} & 0.1549\PM{0.0000}
& 0.8531\PM{0.0000} & 0.1841\PM{0.0000} \\
Self-consistency$_{64}$
& 0.8273\PM{0.0000} & 0.1440\PM{0.0000}
& 0.8875\PM{0.0000} & 0.1605\PM{0.0000}
& 0.8410\PM{0.0000} & 0.1900\PM{0.0000}
& 0.8517\PM{0.0000} & 0.1583\PM{0.0000}
& 0.8505\PM{0.0000} & 0.1849\PM{0.0000} \\
Probe + WV$_{64}$
& 0.8346\PM{0.0030} & 0.1370\PM{0.0009}
& 0.9173\PM{0.0026} & 0.1466\PM{0.0014}
& 0.8844\PM{0.0050} & 0.1649\PM{0.0033}
& 0.9216\PM{0.0066} & 0.1217\PM{0.0050}
& 0.8830\PM{0.0029} & 0.1764\PM{0.0012} \\
\gls{msv} 1 + WV$_{64}$
& 0.8364\PM{0.0019} & 0.1380\PM{0.0005}
& 0.9130\PM{0.0014} & 0.1499\PM{0.0009}
& 0.8829\PM{0.0034} & 0.1682\PM{0.0023}
& 0.9168\PM{0.0031} & 0.1264\PM{0.0018}
& 0.8788\PM{0.0012} & 0.1794\PM{0.0008} \\
PRM + WV$_{64}$
& 0.8274\PM{0.0000} & 0.1418\PM{0.0000}
& 0.8997\PM{0.0000} & 0.1546\PM{0.0000}
& 0.8576\PM{0.0000} & 0.1834\PM{0.0000}
& 0.8784\PM{0.0000} & 0.1473\PM{0.0000}
& 0.8613\PM{0.0000} & 0.1827\PM{0.0000} \\
\gls{msv} 64
& \textbf{0.9049}\PM{0.0162} & \textbf{0.1261}\PM{0.0036}
& \textbf{0.9546}\PM{0.0026} & \textbf{0.0835}\PM{0.0074}
& \textbf{0.9290}\PM{0.0089} & \textbf{0.1267}\PM{0.0039}
& \textbf{0.9822}\PM{0.0003} & \textbf{0.0387}\PM{0.0066}
& \textbf{0.9286}\PM{0.0034} & \textbf{0.1041}\PM{0.0085} \\
\bottomrule
\end{tabular}
}
\end{table}

\begin{table}[t]
\caption{ECE ($\downarrow$) and NLL ($\downarrow$) of verifiers on terminal answers.}
\label{tab:ta_ece_nll}
\setlength{\tabcolsep}{4pt}
\renewcommand{\arraystretch}{0.95}
\centering
\resizebox{\textwidth}{!}{
\begin{tabular}{lrrrrrrrrrr}
\toprule
& \multicolumn{2}{c}{\textbf{MATH}}
& \multicolumn{2}{c}{\textbf{OlympiadBench}}
& \multicolumn{2}{c}{\textbf{AMC12}}
& \multicolumn{2}{c}{\textbf{AIME}}
& \multicolumn{2}{c}{\textbf{Omni\text{-}MATH}} \\
\cmidrule(lr){2-3}\cmidrule(lr){4-5}\cmidrule(lr){6-7}\cmidrule(lr){8-9}\cmidrule(lr){10-11}
\multicolumn{1}{c}{\textbf{Method}}
& \multicolumn{1}{c}{\textbf{ECE $\downarrow$}} & \multicolumn{1}{c}{\textbf{NLL $\downarrow$}}
& \multicolumn{1}{c}{\textbf{ECE $\downarrow$}} & \multicolumn{1}{c}{\textbf{NLL $\downarrow$}}
& \multicolumn{1}{c}{\textbf{ECE $\downarrow$}} & \multicolumn{1}{c}{\textbf{NLL $\downarrow$}}
& \multicolumn{1}{c}{\textbf{ECE $\downarrow$}} & \multicolumn{1}{c}{\textbf{NLL $\downarrow$}}
& \multicolumn{1}{c}{\textbf{ECE $\downarrow$}} & \multicolumn{1}{c}{\textbf{NLL $\downarrow$}} \\
\midrule
Token Probs
& 0.1855\PM{0.0000} & 0.9190\PM{0.0000}
& 0.4124\PM{0.0000} & 1.1444\PM{0.0000}
& 0.3935\PM{0.0000} & 1.1384\PM{0.0000}
& 0.4198\PM{0.0000} & 0.8669\PM{0.0000}
& 0.5389\PM{0.0000} & 1.2505\PM{0.0000} \\
Probe
& 0.0726\PM{0.0401} & 0.4391\PM{0.0485}
& 0.0700\PM{0.0517} & 0.4172\PM{0.0437}
& 0.1064\PM{0.0581} & 0.4680\PM{0.0730}
& 0.1024\PM{0.0483} & 0.3115\PM{0.0488}
& 0.1206\PM{0.0589} & 0.4394\PM{0.0817} \\
\gls{msv} 1
& 0.0849\PM{0.0190} & 0.4787\PM{0.0291}
& 0.0806\PM{0.0232} & 0.4363\PM{0.0213}
& 0.1279\PM{0.0304} & 0.4913\PM{0.0364}
& 0.1043\PM{0.0250} & 0.2993\PM{0.0221}
& 0.1446\PM{0.0288} & 0.4881\PM{0.0345} \\
PRM
& 0.1817\PM{0.0000} & 0.5349\PM{0.0000}
& 0.1834\PM{0.0000} & 0.5231\PM{0.0000}
& 0.1474\PM{0.0000} & 0.5484\PM{0.0000}
& 0.2468\PM{0.0000} & 0.4773\PM{0.0000}
& 0.2160\PM{0.0000} & 0.5010\PM{0.0000} \\
\midrule
Token Probs + WV$_4$
& 0.1615\PM{0.0000} & 1.1805\PM{0.0000}
& 0.2416\PM{0.0000} & 1.2018\PM{0.0000}
& 0.2616\PM{0.0000} & 1.4551\PM{0.0000}
& 0.2677\PM{0.0000} & 1.1803\PM{0.0000}
& 0.3306\PM{0.0000} & 1.4633\PM{0.0000} \\
Self-consistency$_4$
& 0.1611\PM{0.0000} & 1.1823\PM{0.0000}
& 0.2408\PM{0.0000} & 1.2054\PM{0.0000}
& 0.2601\PM{0.0000} & 1.4620\PM{0.0000}
& 0.2664\PM{0.0000} & 1.1889\PM{0.0000}
& 0.3300\PM{0.0000} & 1.4647\PM{0.0000} \\
Probe + WV$_4$
& 0.1655\PM{0.0005} & 1.1793\PM{0.0029}
& 0.2457\PM{0.0007} & 1.2096\PM{0.0066}
& 0.2633\PM{0.0008} & 1.4417\PM{0.0034}
& 0.2703\PM{0.0009} & 1.1423\PM{0.0054}
& 0.3313\PM{0.0003} & 1.4971\PM{0.0116} \\
\gls{msv} 1 + WV$_4$
& 0.1654\PM{0.0004} & 1.1819\PM{0.0031}
& 0.2461\PM{0.0003} & 1.2106\PM{0.0058}
& 0.2641\PM{0.0004} & 1.4449\PM{0.0056}
& 0.2714\PM{0.0006} & 1.1412\PM{0.0020}
& 0.3321\PM{0.0003} & 1.4995\PM{0.0094} \\
PRM + WV$_4$
& 0.1635\PM{0.0000} & 1.1743\PM{0.0000}
& 0.2429\PM{0.0000} & 1.1920\PM{0.0000}
& 0.2627\PM{0.0000} & 1.4440\PM{0.0000}
& 0.2743\PM{0.0000} & 1.1635\PM{0.0000}
& 0.3318\PM{0.0000} & 1.4591\PM{0.0000} \\
\gls{msv} 4
& \textbf{0.1065}\PM{0.0177} & \textbf{0.4677}\PM{0.0362}
& \textbf{0.0821}\PM{0.0267} & \textbf{0.3663}\PM{0.0206}
& \textbf{0.1306}\PM{0.0357} & \textbf{0.4793}\PM{0.0424}
& \textbf{0.0908}\PM{0.0374} & \textbf{0.2472}\PM{0.0384}
& \textbf{0.1265}\PM{0.0309} & \textbf{0.4424}\PM{0.0423} \\
\midrule
Token Probs + WV$_{16}$
& 0.1212\PM{0.0000} & 0.9249\PM{0.0000}
& 0.1566\PM{0.0000} & 0.6566\PM{0.0000}
& 0.1752\PM{0.0000} & 0.8604\PM{0.0000}
& 0.1545\PM{0.0000} & 0.6618\PM{0.0000}
& 0.2235\PM{0.0000} & 0.8289\PM{0.0000} \\
Self-consistency$_{16}$
& 0.1204\PM{0.0000} & 0.9264\PM{0.0000}
& 0.1551\PM{0.0000} & 0.6605\PM{0.0000}
& 0.1737\PM{0.0000} & 0.8673\PM{0.0000}
& 0.1517\PM{0.0000} & 0.6702\PM{0.0000}
& 0.2226\PM{0.0000} & 0.8306\PM{0.0000} \\
Probe + WV$_{16}$
& 0.1293\PM{0.0009} & 0.9282\PM{0.0023}
& 0.1644\PM{0.0015} & 0.6382\PM{0.0023}
& 0.1800\PM{0.0017} & 0.8120\PM{0.0069}
& 0.1579\PM{0.0032} & 0.5808\PM{0.0108}
& 0.2256\PM{0.0010} & 0.8281\PM{0.0047} \\
\gls{msv} 1 + WV$_{16}$
& 0.1290\PM{0.0007} & 0.9299\PM{0.0029}
& 0.1655\PM{0.0005} & 0.6443\PM{0.0030}
& 0.1824\PM{0.0004} & 0.8198\PM{0.0064}
& 0.1611\PM{0.0012} & 0.5922\PM{0.0033}
& 0.2275\PM{0.0005} & 0.8366\PM{0.0053} \\
PRM + WV$_{16}$
& 0.1257\PM{0.0000} & 0.9235\PM{0.0000}
& 0.1597\PM{0.0000} & 0.6439\PM{0.0000}
& 0.1798\PM{0.0000} & 0.8497\PM{0.0000}
& 0.1589\PM{0.0000} & 0.6419\PM{0.0000}
& 0.2264\PM{0.0000} & 0.8255\PM{0.0000} \\
\gls{msv} 16
& \textbf{0.1127}\PM{0.0108} & \textbf{0.4615}\PM{0.0291}
& \textbf{0.0869}\PM{0.0197} & \textbf{0.3289}\PM{0.0200}
& \textbf{0.1317}\PM{0.0197} & \textbf{0.4724}\PM{0.0414}
& \textbf{0.0844}\PM{0.0234} & \textbf{0.2079}\PM{0.0290}
& \textbf{0.1268}\PM{0.0205} & \textbf{0.4385}\PM{0.0411} \\
\midrule
Token Probs + WV$_{64}$
& 0.1158\PM{0.0000} & 0.7592\PM{0.0000}
& 0.1346\PM{0.0000} & 0.5194\PM{0.0000}
& 0.1549\PM{0.0000} & 0.7096\PM{0.0000}
& 0.1254\PM{0.0000} & 0.5219\PM{0.0000}
& 0.1970\PM{0.0000} & 0.6686\PM{0.0000} \\
Self-consistency$_{64}$
& \textbf{0.1149}\PM{0.0000} & 0.7598\PM{0.0000}
& 0.1329\PM{0.0000} & 0.5231\PM{0.0000}
& 0.1531\PM{0.0000} & 0.7161\PM{0.0000}
& 0.1221\PM{0.0000} & 0.5301\PM{0.0000}
& 0.1959\PM{0.0000} & 0.6692\PM{0.0000} \\
Probe + WV$_{64}$
& 0.1205\PM{0.0011} & 0.7788\PM{0.0038}
& 0.1437\PM{0.0018} & 0.4894\PM{0.0039}
& 0.1611\PM{0.0026} & 0.6564\PM{0.0086}
& 0.1288\PM{0.0037} & 0.4265\PM{0.0148}
& 0.1999\PM{0.0014} & 0.6564\PM{0.0032} \\
\gls{msv} 1 + WV$_{64}$
& 0.1202\PM{0.0008} & 0.7732\PM{0.0032}
& 0.1451\PM{0.0007} & 0.5024\PM{0.0022}
& 0.1664\PM{0.0034} & 0.6646\PM{0.0065}
& 0.1347\PM{0.0039} & 0.4444\PM{0.0042}
& 0.2020\PM{0.0007} & 0.6662\PM{0.0044} \\
PRM + WV$_{64}$
& 0.1161\PM{0.0000} & 0.7661\PM{0.0000}
& 0.1382\PM{0.0000} & 0.5060\PM{0.0000}
& 0.1601\PM{0.0000} & 0.7013\PM{0.0000}
& 0.1305\PM{0.0000} & 0.5014\PM{0.0000}
& 0.2004\PM{0.0000} & 0.6658\PM{0.0000} \\
\gls{msv} 64
& 0.1204\PM{0.0109} & \textbf{0.5443}\PM{0.0501}
& \textbf{0.0556}\PM{0.0135} & \textbf{0.2904}\PM{0.0230}
& \textbf{0.1085}\PM{0.0122} & \textbf{0.4986}\PM{0.0234}
& \textbf{0.0350}\PM{0.0166} & \textbf{0.1501}\PM{0.0213}
& \textbf{0.0867}\PM{0.0106} & \textbf{0.3705}\PM{0.0394} \\
\bottomrule
\end{tabular}
}
\end{table}

%% file: table/ta_bon.tex
\begin{table}[t]
\caption{Best-of-N accuracy ($\uparrow$), with verifiers at $N = 1, 4, 16, 64$.}
\label{tab:acc_bon_only_reordered_terminal}
\setlength{\tabcolsep}{5pt}
\renewcommand{\arraystretch}{0.9}
\centering
\resizebox{\textwidth}{!}{
\begin{tabular}{llccccc}
\toprule
\textit{\textbf{N}}& \textbf{Method} & \textbf{MATH} & \textbf{OlympiadBench} & \textbf{AMC12} & \textbf{AIME} & \textbf{Omni\text{-}MATH} \\
\midrule
\multirow{1}{*}{$1$}
& ---                 & 0.6755\PM{0.0000} & 0.4105\PM{0.0000} & 0.3965\PM{0.0000} & 0.2748\PM{0.0000} & 0.2432\PM{0.0000} \\
\midrule
\multirow{10}{*}{$4$}
& Self-consistency$_4$    & 0.7049\PM{0.0000} & 0.4484\PM{0.0000} & 0.4286\PM{0.0000} & 0.3251\PM{0.0000} & 0.2633\PM{0.0000} \\
& Probe               & 0.7197\PM{0.0006} & 0.4725\PM{0.0008} & 0.4861\PM{0.0041} & 0.3771\PM{0.0012} & 0.2885\PM{0.0005} \\
& \gls{msv} 1         & 0.7155\PM{0.0024} & 0.4681\PM{0.0011} & 0.4854\PM{0.0043} & 0.3758\PM{0.0008} & 0.2878\PM{0.0018} \\
& PRM                 & 0.7203\PM{0.0000} & 0.4741\PM{0.0000} & \textbf{0.4879}\PM{0.0000} & 0.3744\PM{0.0000} & \textbf{0.2927}\PM{0.0000} \\
& Probe + WV$_4$          & \textbf{0.7250}\PM{0.0009} & \textbf{0.4770}\PM{0.0003} & 0.4856\PM{0.0030} & 0.3847\PM{0.0010} & 0.2910\PM{0.0009} \\
& \gls{msv} 1 + WV$_4$    & 0.7224\PM{0.0011} & 0.4755\PM{0.0014} & 0.4845\PM{0.0030} & \textbf{0.3852}\PM{0.0008} & 0.2900\PM{0.0012} \\
& PRM + WV$_4$            & 0.7151\PM{0.0000} & 0.4655\PM{0.0000} & 0.4599\PM{0.0000} & 0.3570\PM{0.0000} & 0.2783\PM{0.0000} \\
& \gls{msv} 4         & 0.7219\PM{0.0012} & 0.4747\PM{0.0021} & 0.4837\PM{0.0020} & 0.3811\PM{0.0018} & 0.2891\PM{0.0015} \\
\midrule
\multirow{10}{*}{$16$}
& Self-consistency$_{16}$    & 0.7210\PM{0.0000} & 0.4676\PM{0.0000} & 0.4478\PM{0.0000} & 0.3566\PM{0.0000} & 0.2801\PM{0.0000} \\
& Probe               & 0.7345\PM{0.0014} & 0.5039\PM{0.0023} & 0.5198\PM{0.0037} & 0.4470\PM{0.0025} & 0.3176\PM{0.0020} \\
& \gls{msv} 1         & 0.7238\PM{0.0031} & 0.4895\PM{0.0037} & 0.5231\PM{0.0035} & 0.4325\PM{0.0050} & 0.3102\PM{0.0023} \\
& PRM                 & 0.7383\PM{0.0000} & 0.5128\PM{0.0000} & 0.5280\PM{0.0000} & 0.4475\PM{0.0000} & \textbf{0.3231}\PM{0.0000} \\
& Probe + WV$_{16}$          & 0.7450\PM{0.0028} & 0.5087\PM{0.0019} & 0.5254\PM{0.0072} & 0.4481\PM{0.0096} & 0.3162\PM{0.0037} \\
& \gls{msv} 1 + WV$_{16}$    & 0.7440\PM{0.0025} & 0.5104\PM{0.0008} & 0.5235\PM{0.0055} & 0.4460\PM{0.0076} & 0.3133\PM{0.0039} \\
& PRM + WV$_{16}$            & 0.7266\PM{0.0000} & 0.4860\PM{0.0000} & 0.4701\PM{0.0000} & 0.3817\PM{0.0000} & 0.2879\PM{0.0000} \\
& \gls{msv} 16        & \textbf{0.7472}\PM{0.0009} & \textbf{0.5154}\PM{0.0037} & \textbf{0.5399}\PM{0.0061} & \textbf{0.4615}\PM{0.0046} & 0.3198\PM{0.0032} \\
\midrule
\multirow{10}{*}{$64$}
& Self-consistency$_{64}$    & 0.7321\PM{0.0000} & 0.4844\PM{0.0000} & 0.4403\PM{0.0000} & 0.3594\PM{0.0000} & 0.2790\PM{0.0000} \\
& Probe               & 0.7429\PM{0.0030} & 0.5112\PM{0.0062} & 0.5269\PM{0.0154} & 0.4897\PM{0.0098} & 0.3335\PM{0.0077} \\
& \gls{msv} 1         & 0.7246\PM{0.0058} & 0.4629\PM{0.0110} & 0.5164\PM{0.0119} & 0.4621\PM{0.0140} & 0.3179\PM{0.0061} \\
& PRM                 & 0.7388\PM{0.0000} & 0.5223\PM{0.0000} & 0.5522\PM{0.0000} & 0.4688\PM{0.0000} & 0.3393\PM{0.0000} \\
& Probe + WV$_{64}$          & 0.7500\PM{0.0032} & 0.5223\PM{0.0058} & 0.5388\PM{0.0073} & 0.4790\PM{0.0067} & 0.3210\PM{0.0017} \\
& \gls{msv} 1 + WV$_{64}$    & 0.7442\PM{0.0033} & 0.5210\PM{0.0030} & 0.5313\PM{0.0087} & 0.4714\PM{0.0052} & 0.3179\PM{0.0023} \\
& PRM + WV$_{64}$            & 0.7344\PM{0.0000} & 0.4933\PM{0.0000} & 0.4851\PM{0.0000} & 0.3884\PM{0.0000} & 0.2969\PM{0.0000} \\
& \gls{msv} 64        & \textbf{0.7607}\PM{0.0033} & \textbf{0.5366}\PM{0.0064} & \textbf{0.5866}\PM{0.0138} & \textbf{0.5098}\PM{0.0073} & \textbf{0.3540}\PM{0.0023} \\
\bottomrule
\end{tabular}
}
\end{table}

\begin{table}[t]
\caption{ECE ($\downarrow$) and Brier score ($\downarrow$) of confidence output on best-of-$N$ answers, with verifiers at $N = 4, 16, 64$.}
\label{tab:calib_bon_only_reordered_terminal}
\setlength{\tabcolsep}{3pt}
\renewcommand{\arraystretch}{0.9}
\centering
\resizebox{\textwidth}{!}{
\begin{tabular}{llcccccccccc}
\toprule
&  & \multicolumn{2}{c}{\textbf{MATH}} & \multicolumn{2}{c}{\textbf{OlympiadBench}} & \multicolumn{2}{c}{\textbf{AMC12}} & \multicolumn{2}{c}{\textbf{AIME}} & \multicolumn{2}{c}{\textbf{Omni\text{-}MATH}} \\
\cmidrule(lr){3-4}\cmidrule(lr){5-6}\cmidrule(lr){7-8}\cmidrule(lr){9-10}\cmidrule(lr){11-12}
\multicolumn{1}{c}{\textbf{\textit{N}}} & \multicolumn{1}{c}{\textbf{Method}}
& \multicolumn{1}{c}{\textbf{ECE $\downarrow$}} & \multicolumn{1}{c}{\textbf{Brier $\downarrow$}}
& \multicolumn{1}{c}{\textbf{ECE $\downarrow$}} & \multicolumn{1}{c}{\textbf{Brier $\downarrow$}}
& \multicolumn{1}{c}{\textbf{ECE $\downarrow$}} & \multicolumn{1}{c}{\textbf{Brier $\downarrow$}}
& \multicolumn{1}{c}{\textbf{ECE $\downarrow$}} & \multicolumn{1}{c}{\textbf{Brier $\downarrow$}}
& \multicolumn{1}{c}{\textbf{ECE $\downarrow$}} & \multicolumn{1}{c}{\textbf{Brier $\downarrow$}} \\
\midrule
\multirow{10}{*}{$4$}
& Self-consistency$_4$
& 0.1613\PM{0.0000} & 0.3638\PM{0.0000}
& 0.2586\PM{0.0000} & 0.4983\PM{0.0000}
& 0.2843\PM{0.0000} & 0.5504\PM{0.0000}
& 0.2805\PM{0.0000} & 0.5316\PM{0.0000}
& 0.3708\PM{0.0000} & 0.6031\PM{0.0000} \\
& PRM
& 0.1308\PM{0.0000} & 0.3283\PM{0.0000}
& 0.2016\PM{0.0000} & 0.3508\PM{0.0000}
& 0.1386\PM{0.0000} & 0.3888\PM{0.0000}
& 0.2363\PM{0.0000} & 0.3276\PM{0.0000}
& 0.2241\PM{0.0000} & 0.3614\PM{0.0000} \\
& PRM + WV$_4$
& 0.1607\PM{0.0000} & 0.3645\PM{0.0000}
& 0.2554\PM{0.0000} & 0.5029\PM{0.0000}
& 0.2687\PM{0.0000} & 0.5582\PM{0.0000}
& 0.2785\PM{0.0000} & 0.5357\PM{0.0000}
& 0.3724\PM{0.0000} & 0.6174\PM{0.0000} \\
& Probe
& 0.1168\PM{0.0448} & \textbf{0.3001}\PM{0.0341}
& 0.1510\PM{0.0689} & 0.3376\PM{0.0653}
& 0.1801\PM{0.0639} & 0.3806\PM{0.0666}
& 0.1703\PM{0.0579} & 0.2587\PM{0.0594}
& 0.2298\PM{0.0777} & 0.3997\PM{0.0995} \\
& \gls{msv} 1
& 0.1316\PM{0.0206} & 0.3316\PM{0.0176}
& 0.1675\PM{0.0276} & 0.3624\PM{0.0257}
& 0.1985\PM{0.0293} & 0.3927\PM{0.0311}
& 0.1661\PM{0.0269} & 0.2459\PM{0.0242}
& 0.2514\PM{0.0319} & 0.4372\PM{0.0374} \\
& Probe + WV$_4$
& 0.1797\PM{0.0039} & 0.3856\PM{0.0043}
& 0.3066\PM{0.0089} & 0.5660\PM{0.0129}
& 0.3050\PM{0.0073} & 0.6019\PM{0.0122}
& 0.3204\PM{0.0133} & 0.5777\PM{0.0180}
& 0.4396\PM{0.0109} & 0.7359\PM{0.0210} \\
& \gls{msv} 1 + WV$_4$
& 0.1815\PM{0.0034} & 0.3871\PM{0.0036}
& 0.3044\PM{0.0066} & 0.5642\PM{0.0105}
& 0.3014\PM{0.0063} & 0.5945\PM{0.0065}
& 0.3149\PM{0.0079} & 0.5680\PM{0.0066}
& 0.4354\PM{0.0091} & 0.7306\PM{0.0156} \\
& \gls{msv} 4
& \textbf{0.1133}\PM{0.0187} & 0.3010\PM{0.0118}
& \textbf{0.1031}\PM{0.0345} & \textbf{0.2864}\PM{0.0178}
& \textbf{0.1477}\PM{0.0412} & \textbf{0.3477}\PM{0.0268}
& \textbf{0.1019}\PM{0.0489} & \textbf{0.2040}\PM{0.0319}
& \textbf{0.1695}\PM{0.0402} & \textbf{0.3410}\PM{0.0364} \\
\midrule
\multirow{10}{*}{$16$}
& Self-consistency$_{16}$
& 0.1292\PM{0.0000} & 0.3404\PM{0.0000}
& 0.1971\PM{0.0000} & 0.4347\PM{0.0000}
& 0.2235\PM{0.0000} & 0.4840\PM{0.0000}
& 0.1927\PM{0.0000} & 0.4558\PM{0.0000}
& 0.2961\PM{0.0000} & 0.5021\PM{0.0000} \\
& PRM
& \textbf{0.1092}\PM{0.0000} & 0.3163\PM{0.0000}
& 0.1946\PM{0.0000} & 0.3590\PM{0.0000}
& \textbf{0.1162}\PM{0.0000} & 0.3999\PM{0.0000}
& 0.2288\PM{0.0000} & 0.3452\PM{0.0000}
& 0.2227\PM{0.0000} & 0.3955\PM{0.0000} \\
& PRM + WV$_{16}$
& 0.1334\PM{0.0000} & 0.3356\PM{0.0000}
& 0.1908\PM{0.0000} & 0.4329\PM{0.0000}
& 0.2138\PM{0.0000} & 0.4774\PM{0.0000}
& 0.1841\PM{0.0000} & 0.4438\PM{0.0000}
& 0.2986\PM{0.0000} & 0.5036\PM{0.0000} \\
& Probe
& 0.1559\PM{0.0369} & 0.3397\PM{0.0415}
& 0.2294\PM{0.0679} & 0.4428\PM{0.0920}
& 0.2657\PM{0.0607} & 0.5024\PM{0.0907}
& 0.2364\PM{0.0601} & 0.3776\PM{0.0858}
& 0.3296\PM{0.0823} & 0.5635\PM{0.1360} \\
& \gls{msv} 1
& 0.1740\PM{0.0172} & 0.3818\PM{0.0229}
& 0.2519\PM{0.0267} & 0.4830\PM{0.0365}
& 0.2725\PM{0.0268} & 0.5147\PM{0.0381}
& 0.2407\PM{0.0280} & 0.3754\PM{0.0442}
& 0.3542\PM{0.0327} & 0.6170\PM{0.0501} \\
& Probe + WV$_{16}$
& 0.1353\PM{0.0017} & 0.3407\PM{0.0023}
& 0.2073\PM{0.0070} & 0.4405\PM{0.0035}
& 0.1988\PM{0.0046} & 0.4744\PM{0.0030}
& 0.1633\PM{0.0035} & 0.4286\PM{0.0055}
& 0.3195\PM{0.0064} & 0.5521\PM{0.0111} \\
& \gls{msv} 1 + WV$_{16}$
& 0.1356\PM{0.0007} & 0.3452\PM{0.0024}
& 0.2049\PM{0.0053} & 0.4476\PM{0.0042}
& 0.2002\PM{0.0019} & 0.4760\PM{0.0014}
& 0.1659\PM{0.0040} & 0.4352\PM{0.0036}
& 0.3201\PM{0.0056} & 0.5540\PM{0.0093} \\
& \gls{msv} 16
& 0.1160\PM{0.0142} & \textbf{0.2905}\PM{0.0079}
& \textbf{0.1059}\PM{0.0288} & \textbf{0.2859}\PM{0.0204}
& 0.1462\PM{0.0273} & \textbf{0.3516}\PM{0.0216}
& \textbf{0.0876}\PM{0.0301} & \textbf{0.2067}\PM{0.0236}
& \textbf{0.1776}\PM{0.0328} & \textbf{0.3583}\PM{0.0330} \\
\midrule
\multirow{10}{*}{$64$}
& Self-consistency$_{64}$
& 0.1259\PM{0.0000} & 0.3377\PM{0.0000}
& 0.1680\PM{0.0000} & 0.4279\PM{0.0000}
& 0.2196\PM{0.0000} & 0.4502\PM{0.0000}
& 0.1733\PM{0.0000} & 0.4341\PM{0.0000}
& 0.2830\PM{0.0000} & 0.4749\PM{0.0000} \\
& PRM
& \textbf{0.0884}\PM{0.0000} & 0.3200\PM{0.0000}
& 0.1667\PM{0.0000} & 0.3719\PM{0.0000}
& \textbf{0.1174}\PM{0.0000} & 0.4124\PM{0.0000}
& 0.2167\PM{0.0000} & 0.3627\PM{0.0000}
& 0.2326\PM{0.0000} & 0.4304\PM{0.0000} \\
& PRM + WV$_{64}$
& 0.1269\PM{0.0000} & 0.3303\PM{0.0000}
& 0.1694\PM{0.0000} & 0.4145\PM{0.0000}
& 0.1885\PM{0.0000} & 0.4678\PM{0.0000}
& 0.1606\PM{0.0000} & 0.4253\PM{0.0000}
& 0.2757\PM{0.0000} & 0.4885\PM{0.0000} \\
& Probe
& 0.1848\PM{0.0276} & 0.3802\PM{0.0400}
& 0.3048\PM{0.0552} & 0.5520\PM{0.0995}
& 0.3288\PM{0.0573} & 0.6132\PM{0.1003}
& 0.3013\PM{0.0588} & 0.5159\PM{0.1006}
& 0.4185\PM{0.0796} & 0.7348\PM{0.1555} \\
& \gls{msv} 1
& 0.2128\PM{0.0143} & 0.4419\PM{0.0240}
& 0.3555\PM{0.0312} & 0.6453\PM{0.0520}
& 0.3520\PM{0.0174} & 0.6636\PM{0.0295}
& 0.3162\PM{0.0348} & 0.5324\PM{0.0653}
& 0.4504\PM{0.0303} & 0.8118\PM{0.0537} \\
& Probe + WV$_{64}$
& 0.1259\PM{0.0029} & 0.3305\PM{0.0019}
& 0.1736\PM{0.0020} & 0.4102\PM{0.0041}
& 0.1823\PM{0.0155} & 0.4611\PM{0.0072}
& 0.1222\PM{0.0168} & 0.4165\PM{0.0138}
& 0.2888\PM{0.0049} & 0.5115\PM{0.0042} \\
& \gls{msv} 1 + WV$_{64}$
& 0.1300\PM{0.0016} & 0.3329\PM{0.0017}
& 0.1769\PM{0.0059} & 0.4219\PM{0.0016}
& 0.1929\PM{0.0086} & 0.4586\PM{0.0038}
& 0.1302\PM{0.0100} & 0.4232\PM{0.0069}
& 0.2920\PM{0.0033} & 0.5175\PM{0.0060} \\
& \gls{msv} 64
& 0.1323\PM{0.0147} & \textbf{0.2974}\PM{0.0080}
& \textbf{0.0803}\PM{0.0281} & \textbf{0.2613}\PM{0.0315}
& 0.1731\PM{0.0207} & \textbf{0.4103}\PM{0.0187}
& \textbf{0.0749}\PM{0.0500} & \textbf{0.2351}\PM{0.0542}
& \textbf{0.1303}\PM{0.0273} & \textbf{0.3434}\PM{0.0401} \\
\bottomrule
\end{tabular}
}
\end{table}

%% file: table/sa_standard.tex
\begin{table}[t]
\caption{AUROC ($\uparrow$) and Brier score ($\downarrow$) of verifiers on streaming answers.}
\label{tab:sa_standard_auroc_brier}
\setlength{\tabcolsep}{3pt}
\renewcommand{\arraystretch}{0.9}
\centering
\resizebox{\textwidth}{!}{
\begin{tabular}{lrrrrrrrrrr}
\toprule
& \multicolumn{2}{c}{\textbf{MATH}}
& \multicolumn{2}{c}{\textbf{OlympiadBench}}
& \multicolumn{2}{c}{\textbf{AMC12}}
& \multicolumn{2}{c}{\textbf{AIME}}
& \multicolumn{2}{c}{\textbf{Omni\text{-}MATH}} \\
\cmidrule(lr){2-3}\cmidrule(lr){4-5}\cmidrule(lr){6-7}\cmidrule(lr){8-9}\cmidrule(lr){10-11}
\multicolumn{1}{c}{\textbf{Method}}
& \multicolumn{1}{c}{\textbf{AUROC $\uparrow$}} & \multicolumn{1}{c}{\textbf{Brier $\downarrow$}}
& \multicolumn{1}{c}{\textbf{AUROC $\uparrow$}} & \multicolumn{1}{c}{\textbf{Brier $\downarrow$}}
& \multicolumn{1}{c}{\textbf{AUROC $\uparrow$}} & \multicolumn{1}{c}{\textbf{Brier $\downarrow$}}
& \multicolumn{1}{c}{\textbf{AUROC $\uparrow$}} & \multicolumn{1}{c}{\textbf{Brier $\downarrow$}}
& \multicolumn{1}{c}{\textbf{AUROC $\uparrow$}} & \multicolumn{1}{c}{\textbf{Brier $\downarrow$}} \\
\midrule
Token Probs
& 0.7910\PM{0.0000} & 0.3399\PM{0.0000}
& 0.8091\PM{0.0000} & 0.4266\PM{0.0000}
& 0.7856\PM{0.0000} & 0.3979\PM{0.0000}
& 0.8858\PM{0.0000} & 0.3217\PM{0.0000}
& 0.7974\PM{0.0000} & 0.4746\PM{0.0000} \\
PRM
& \textbf{0.8692}\PM{0.0000} & 0.1902\PM{0.0000}
& \textbf{0.8542}\PM{0.0000} & 0.1607\PM{0.0000}
& \textbf{0.8658}\PM{0.0000} & 0.1621\PM{0.0000}
& \textbf{0.9309}\PM{0.0000} & 0.1185\PM{0.0000}
& 0.8227\PM{0.0000} & 0.1497\PM{0.0000} \\
Probe
& 0.8690\PM{0.0135} & \textbf{0.1592}\PM{0.0101}
& 0.8317\PM{0.0089} & 0.1478\PM{0.0232}
& 0.8621\PM{0.0108} & 0.1413\PM{0.0195}
& 0.9120\PM{0.0141} & 0.0796\PM{0.0105}
& \textbf{0.8229}\PM{0.0142} & 0.1527\PM{0.0242} \\
\gls{msv} 1
& 0.8443\PM{0.0223} & 0.1601\PM{0.0125}
& 0.8169\PM{0.0135} & \textbf{0.1425}\PM{0.0076}
& 0.8546\PM{0.0104} & \textbf{0.1343}\PM{0.0028}
& 0.9056\PM{0.0076} & \textbf{0.0642}\PM{0.0028}
& 0.7951\PM{0.0125} & \textbf{0.1381}\PM{0.0084} \\
\midrule
Token Probs + WV$_4$
& 0.7743\PM{0.0000} & 0.3266\PM{0.0000}
& 0.7837\PM{0.0000} & 0.4051\PM{0.0000}
& 0.7596\PM{0.0000} & 0.3775\PM{0.0000}
& 0.8601\PM{0.0000} & 0.2990\PM{0.0000}
& 0.7757\PM{0.0000} & 0.4522\PM{0.0000} \\
PRM + WV$_4$
& 0.8533\PM{0.0000} & 0.2006\PM{0.0000}
& 0.8142\PM{0.0000} & 0.1678\PM{0.0000}
& 0.8296\PM{0.0000} & 0.1689\PM{0.0000}
& 0.8646\PM{0.0000} & 0.1241\PM{0.0000}
& 0.7975\PM{0.0000} & 0.1494\PM{0.0000} \\
Probe + WV$_4$
& 0.8528\PM{0.0027} & 0.1619\PM{0.0044}
& 0.8104\PM{0.0037} & 0.1430\PM{0.0095}
& 0.8219\PM{0.0095} & 0.1395\PM{0.0106}
& 0.8702\PM{0.0206} & 0.0842\PM{0.0067}
& 0.8009\PM{0.0079} & 0.1331\PM{0.0200} \\
\gls{msv} 1 + WV$_4$
& 0.8367\PM{0.0150} & 0.1696\PM{0.0113}
& 0.7974\PM{0.0058} & 0.1420\PM{0.0029}
& 0.8427\PM{0.0100} & 0.1330\PM{0.0059}
& 0.8826\PM{0.0068} & 0.0763\PM{0.0020}
& 0.7699\PM{0.0098} & 0.1228\PM{0.0081} \\
\gls{msv} 4
& \textbf{0.8917}\PM{0.0133} & \textbf{0.1324}\PM{0.0034}
& \textbf{0.8683}\PM{0.0169} & \textbf{0.1197}\PM{0.0094}
& \textbf{0.9002}\PM{0.0118} & \textbf{0.1068}\PM{0.0033}
& \textbf{0.9312}\PM{0.0101} & \textbf{0.0472}\PM{0.0053}
& \textbf{0.8327}\PM{0.0127} & \textbf{0.1160}\PM{0.0088} \\
\midrule
Token Probs + WV$_{16}$
& 0.7681\PM{0.0000} & 0.3242\PM{0.0000}
& 0.7804\PM{0.0000} & 0.4029\PM{0.0000}
& 0.7550\PM{0.0000} & 0.3750\PM{0.0000}
& 0.8604\PM{0.0000} & 0.2967\PM{0.0000}
& 0.7724\PM{0.0000} & 0.4507\PM{0.0000} \\
PRM + WV$_{16}$
& 0.8430\PM{0.0000} & 0.2075\PM{0.0000}
& 0.8018\PM{0.0000} & 0.1715\PM{0.0000}
& 0.8156\PM{0.0000} & 0.1725\PM{0.0000}
& 0.8497\PM{0.0000} & 0.1266\PM{0.0000}
& 0.7901\PM{0.0000} & 0.1508\PM{0.0000} \\
Probe + WV$_{16}$
& 0.8414\PM{0.0015} & 0.1666\PM{0.0024}
& 0.8013\PM{0.0037} & 0.1453\PM{0.0085}
& 0.8113\PM{0.0111} & 0.1418\PM{0.0088}
& 0.8609\PM{0.0214} & 0.0873\PM{0.0060}
& 0.7966\PM{0.0076} & 0.1315\PM{0.0193} \\
\gls{msv} 1 + WV$_{16}$
& 0.8255\PM{0.0150} & 0.1761\PM{0.0118}
& 0.7889\PM{0.0056} & 0.1450\PM{0.0027}
& 0.8341\PM{0.0102} & 0.1367\PM{0.0059}
& 0.8749\PM{0.0068} & 0.0804\PM{0.0020}
& 0.7634\PM{0.0099} & 0.1217\PM{0.0078} \\
\gls{msv} 16
& \textbf{0.9037}\PM{0.0201} & \textbf{0.1104}\PM{0.0105}
& \textbf{0.8946}\PM{0.0229} & \textbf{0.1054}\PM{0.0076}
& \textbf{0.9105}\PM{0.0115} & \textbf{0.1041}\PM{0.0049}
& \textbf{0.9532}\PM{0.0069} & \textbf{0.0375}\PM{0.0029}
& \textbf{0.8591}\PM{0.0029} & \textbf{0.1100}\PM{0.0063} \\
\midrule
Token Probs + WV$_{64}$
& 0.7676\PM{0.0000} & 0.3238\PM{0.0000}
& 0.7798\PM{0.0000} & 0.4026\PM{0.0000}
& 0.7527\PM{0.0000} & 0.3749\PM{0.0000}
& 0.8592\PM{0.0000} & 0.2966\PM{0.0000}
& 0.7718\PM{0.0000} & 0.4504\PM{0.0000} \\
PRM + WV$_{64}$
& 0.8385\PM{0.0000} & 0.2098\PM{0.0000}
& 0.7990\PM{0.0000} & 0.1723\PM{0.0000}
& 0.8121\PM{0.0000} & 0.1733\PM{0.0000}
& 0.8451\PM{0.0000} & 0.1272\PM{0.0000}
& 0.7881\PM{0.0000} & 0.1512\PM{0.0000} \\
Probe + WV$_{64}$
& 0.8388\PM{0.0015} & 0.1679\PM{0.0019}
& 0.7993\PM{0.0038} & 0.1457\PM{0.0082}
& 0.8089\PM{0.0112} & 0.1421\PM{0.0085}
& 0.8588\PM{0.0214} & 0.0878\PM{0.0059}
& 0.7963\PM{0.0076} & 0.1310\PM{0.0191} \\
\gls{msv} 1 + WV$_{64}$
& 0.8225\PM{0.0152} & 0.1778\PM{0.0119}
& 0.7871\PM{0.0057} & 0.1455\PM{0.0027}
& 0.8323\PM{0.0105} & 0.1375\PM{0.0060}
& 0.8731\PM{0.0069} & 0.0812\PM{0.0020}
& 0.7622\PM{0.0101} & 0.1214\PM{0.0078} \\
\gls{msv} 64
& \textbf{0.9215}\PM{0.0125} & \textbf{0.1137}\PM{0.0022}
& \textbf{0.8971}\PM{0.0136} & \textbf{0.0956}\PM{0.0118}
& \textbf{0.8917}\PM{0.0034} & \textbf{0.1154}\PM{0.0033}
& \textbf{0.9662}\PM{0.0016} & \textbf{0.0309}\PM{0.0013}
& \textbf{0.8409}\PM{0.0084} & \textbf{0.1205}\PM{0.0051} \\
\bottomrule
\end{tabular}
}
\end{table}

\begin{table}[t]
\caption{ECE ($\downarrow$) and NLL ($\downarrow$) of verifiers on streaming answers.}
\label{tab:sa_standard_ece_nll}
\setlength{\tabcolsep}{3pt}
\renewcommand{\arraystretch}{0.9}
\centering
\resizebox{\textwidth}{!}{
\begin{tabular}{lrrrrrrrrrr}
\toprule
& \multicolumn{2}{c}{\textbf{MATH}}
& \multicolumn{2}{c}{\textbf{OlympiadBench}}
& \multicolumn{2}{c}{\textbf{AMC12}}
& \multicolumn{2}{c}{\textbf{AIME}}
& \multicolumn{2}{c}{\textbf{Omni\text{-}MATH}} \\
\cmidrule(lr){2-3}\cmidrule(lr){4-5}\cmidrule(lr){6-7}\cmidrule(lr){8-9}\cmidrule(lr){10-11}
\multicolumn{1}{c}{\textbf{Method}}
& \multicolumn{1}{c}{\textbf{ECE $\downarrow$}} & \multicolumn{1}{c}{\textbf{NLL $\downarrow$}}
& \multicolumn{1}{c}{\textbf{ECE $\downarrow$}} & \multicolumn{1}{c}{\textbf{NLL $\downarrow$}}
& \multicolumn{1}{c}{\textbf{ECE $\downarrow$}} & \multicolumn{1}{c}{\textbf{NLL $\downarrow$}}
& \multicolumn{1}{c}{\textbf{ECE $\downarrow$}} & \multicolumn{1}{c}{\textbf{NLL $\downarrow$}}
& \multicolumn{1}{c}{\textbf{ECE $\downarrow$}} & \multicolumn{1}{c}{\textbf{NLL $\downarrow$}} \\
\midrule
Token Probs
& 0.3732\PM{0.0000} & 1.1043\PM{0.0000}
& 0.5255\PM{0.0000} & 1.1625\PM{0.0000}
& 0.4944\PM{0.0000} & 1.1466\PM{0.0000}
& 0.4886\PM{0.0000} & 0.8724\PM{0.0000}
& 0.6045\PM{0.0000} & 1.2632\PM{0.0000} \\
PRM
& 0.2060\PM{0.0000} & 0.5667\PM{0.0000}
& 0.1883\PM{0.0000} & 0.5046\PM{0.0000}
& 0.1775\PM{0.0000} & 0.5076\PM{0.0000}
& 0.2456\PM{0.0000} & 0.4127\PM{0.0000}
& 0.2298\PM{0.0000} & 0.4819\PM{0.0000} \\
Probe
& \textbf{0.0934}\PM{0.0416} & \textbf{0.5184}\PM{0.0528}
& 0.1317\PM{0.0382} & 0.5206\PM{0.0914}
& 0.1292\PM{0.0461} & 0.4841\PM{0.0689}
& 0.1160\PM{0.0385} & 0.2847\PM{0.0421}
& 0.1656\PM{0.0470} & 0.5194\PM{0.0945} \\
\gls{msv} 1
& 0.0939\PM{0.0163} & 0.5356\PM{0.0441}
& \textbf{0.0911}\PM{0.0163} & \textbf{0.5055}\PM{0.0208}
& \textbf{0.0862}\PM{0.0164} & \textbf{0.4532}\PM{0.0039}
& \textbf{0.0511}\PM{0.0229} & \textbf{0.2412}\PM{0.0113}
& \textbf{0.1157}\PM{0.0132} & \textbf{0.4974}\PM{0.0232} \\
\midrule
Token Probs + WV$_4$
& 0.3499\PM{0.0000} & 0.9724\PM{0.0000}
& 0.5001\PM{0.0000} & 1.0713\PM{0.0000}
& 0.4680\PM{0.0000} & 1.0312\PM{0.0000}
& 0.4595\PM{0.0000} & 0.8013\PM{0.0000}
& 0.5843\PM{0.0000} & 1.1794\PM{0.0000} \\
PRM + WV$_4$
& 0.1974\PM{0.0000} & 0.5887\PM{0.0000}
& 0.1705\PM{0.0000} & 0.5201\PM{0.0000}
& 0.1690\PM{0.0000} & 0.5224\PM{0.0000}
& 0.2097\PM{0.0000} & 0.4243\PM{0.0000}
& 0.2165\PM{0.0000} & 0.4822\PM{0.0000} \\
Probe + WV$_4$
& \textbf{0.0628}\PM{0.0368} & 0.4975\PM{0.0160}
& 0.0784\PM{0.0407} & 0.4530\PM{0.0219}
& 0.0790\PM{0.0444} & 0.4438\PM{0.0281}
& 0.0830\PM{0.0286} & 0.2985\PM{0.0248}
& 0.1404\PM{0.0546} & 0.4260\PM{0.0553} \\
\gls{msv} 1 + WV$_4$
& 0.0694\PM{0.0249} & 0.5329\PM{0.0369}
& \textbf{0.0521}\PM{0.0087} & 0.4627\PM{0.0083}
& \textbf{0.0396}\PM{0.0171} & 0.4168\PM{0.0163}
& \textbf{0.0240}\PM{0.0113} & 0.2600\PM{0.0072}
& \textbf{0.0843}\PM{0.0154} & \textbf{0.4068}\PM{0.0196} \\
\gls{msv} 4
& 0.0814\PM{0.0133} & \textbf{0.4637}\PM{0.0294}
& 0.0797\PM{0.0143} & \textbf{0.4289}\PM{0.0270}
& 0.0638\PM{0.0093} & \textbf{0.3709}\PM{0.0172}
& 0.0271\PM{0.0120} & \textbf{0.1814}\PM{0.0187}
& 0.0959\PM{0.0116} & 0.4206\PM{0.0287} \\
\midrule
Token Probs + WV$_{16}$
& 0.3441\PM{0.0000} & 0.9451\PM{0.0000}
& 0.4967\PM{0.0000} & 1.0605\PM{0.0000}
& 0.4643\PM{0.0000} & 1.0155\PM{0.0000}
& 0.4566\PM{0.0000} & 0.7951\PM{0.0000}
& 0.5826\PM{0.0000} & 1.1731\PM{0.0000} \\
PRM + WV$_{16}$
& 0.1969\PM{0.0000} & 0.6035\PM{0.0000}
& 0.1700\PM{0.0000} & 0.5280\PM{0.0000}
& 0.1653\PM{0.0000} & 0.5302\PM{0.0000}
& 0.2027\PM{0.0000} & 0.4301\PM{0.0000}
& 0.2143\PM{0.0000} & 0.4855\PM{0.0000} \\
Probe + WV$_{16}$
& \textbf{0.0573}\PM{0.0197} & 0.5039\PM{0.0067}
& 0.0730\PM{0.0375} & 0.4563\PM{0.0191}
& 0.0734\PM{0.0379} & 0.4476\PM{0.0225}
& 0.0854\PM{0.0256} & 0.3080\PM{0.0233}
& 0.1374\PM{0.0550} & 0.4180\PM{0.0514} \\
\gls{msv} 1 + WV$_{16}$
& 0.0709\PM{0.0311} & 0.5404\PM{0.0354}
& \textbf{0.0497}\PM{0.0131} & 0.4651\PM{0.0080}
& \textbf{0.0427}\PM{0.0169} & 0.4228\PM{0.0161}
& \textbf{0.0295}\PM{0.0103} & 0.2705\PM{0.0066}
& \textbf{0.0784}\PM{0.0170} & 0.3969\PM{0.0188} \\
\gls{msv} 16
& 0.0666\PM{0.0211} & \textbf{0.3781}\PM{0.0461}
& 0.0606\PM{0.0166} & \textbf{0.3546}\PM{0.0195}
& 0.0563\PM{0.0108} & \textbf{0.3672}\PM{0.0173}
& 0.0333\PM{0.0146} & \textbf{0.1498}\PM{0.0109}
& 0.0883\PM{0.0122} & \textbf{0.3888}\PM{0.0163} \\
\midrule
Token Probs + WV$_{64}$
& 0.3429\PM{0.0000} & 0.9397\PM{0.0000}
& 0.4962\PM{0.0000} & 1.0584\PM{0.0000}
& 0.4637\PM{0.0000} & 1.0133\PM{0.0000}
& 0.4561\PM{0.0000} & 0.7946\PM{0.0000}
& 0.5823\PM{0.0000} & 1.1714\PM{0.0000} \\
PRM + WV$_{64}$
& 0.1959\PM{0.0000} & 0.6084\PM{0.0000}
& 0.1697\PM{0.0000} & 0.5299\PM{0.0000}
& 0.1677\PM{0.0000} & 0.5321\PM{0.0000}
& 0.2016\PM{0.0000} & 0.4315\PM{0.0000}
& 0.2138\PM{0.0000} & 0.4866\PM{0.0000} \\
Probe + WV$_{64}$
& \textbf{0.0583}\PM{0.0146} & 0.5063\PM{0.0047}
& 0.0728\PM{0.0365} & 0.4570\PM{0.0186}
& 0.0764\PM{0.0338} & 0.4482\PM{0.0215}
& 0.0871\PM{0.0255} & 0.3099\PM{0.0230}
& 0.1368\PM{0.0551} & 0.4161\PM{0.0506} \\
\gls{msv} 1 + WV$_{64}$
& 0.0724\PM{0.0316} & 0.5428\PM{0.0351}
& \textbf{0.0504}\PM{0.0134} & 0.4651\PM{0.0079}
& \textbf{0.0444}\PM{0.0162} & 0.4244\PM{0.0159}
& 0.0312\PM{0.0093} & 0.2727\PM{0.0065}
& \textbf{0.0769}\PM{0.0176} & \textbf{0.3946}\PM{0.0187} \\
\gls{msv} 64
& 0.0682\PM{0.0151} & \textbf{0.3989}\PM{0.0322}
& 0.0661\PM{0.0236} & \textbf{0.3370}\PM{0.0346}
& 0.0995\PM{0.0065} & \textbf{0.4117}\PM{0.0210}
& \textbf{0.0097}\PM{0.0048} & \textbf{0.1202}\PM{0.0027}
& 0.1242\PM{0.0069} & 0.3998\PM{0.0284} \\
\bottomrule
\end{tabular}
}
\end{table}